\documentclass[preprint,showpacs,showkeys,preprintnumbers,amsmath,amssymb,aps]{revtex4}

\usepackage{mathrsfs}
\usepackage{graphicx}
\usepackage{dcolumn}
\usepackage{bm}
\newtheorem{theorem}{Theorem}[section]
\newtheorem{lemma}[theorem]{Lemma}

\newtheorem{remark}[theorem]{Remark}
\newenvironment{proof}[1][Proof.]{\begin{trivlist}
     \item[\hskip \labelsep {\bfseries #1}]}{\end{trivlist}}

\begin{document}

\preprint{APS/123-QED}

\title{Scattering for massive Dirac fields on the Kerr metric}

\author{D.Batic}

\email{batic@itp.phys.ethz.ch}
\affiliation{ 
Institute for Theoretical Physics\\
Swiss Federal Institute of Technology\\
CH-8093 Z\"{u}rich, Switzerland}

\date{\today} 

\begin{abstract}
Starting with the Dirac equation outside the event horizon of a non-extreme Kerr black hole, we develop a time-dependent scattering theory for massive Dirac particles. The explicit computation of the modified wave operators at infinity is done by implementing a time-dependent logarithmic phase shift from the free dynamics to offset the long range term in the full Hamiltonian due to the presence of the gravitational force. Analytical expressions for the wave operators are also given.
\end{abstract}

\pacs{03.65Nk, 04.62.+v, 04.70.Dy}
\keywords{Dirac equation, general relativity, Kerr metric, time-dependent scattering theory}
\maketitle

\section{\label{sec:1}Introduction}
In this paper we develop a time-dependent scattering theory for massive Dirac particles outside the event horizon of a non-extreme Kerr black hole manifold. Such a solution of the vacuum Einstein field equations discovered by Kerr in 1963 describes an asymptotically flat space-time containing nothing but an eternal, rotating black hole, and has been generalized to the case of charged, spinning black holes by Newman et al. in 1965. Although it is not the most general model of the exterior region of a black hole we can analyze theoretically, it represents indeed the most realistic model in astrophysics since in general black holes are embedded in environments that are rich in gas and plasma and, consequently any net charge is neutralized by the ambient plasma (see for instance Misner et al. (1999) p.885).\\
The aim of scattering theory on curved space-times is to provide a detailed description of the asymptotic behavior in time of some field (in our case Dirac fields). A general feature of scattering theories on black hole manifolds is that two asymptotic regions are present. In fact, in the analysis of the propagation of fields outside the event horizon of a black hole we commonly adopt the point of view of an observer static at space-like infinity, who in turns perceives the event horizon as an asymptotic region.\\
There are mainly three motivations for the analysis of our problem. The first one, quite general on its own, aims to provide a deeper insight into the physics of black holes with particular attention to the quantum field theory on curved space-times. The second one is the rigorous mathematical analysis of quantum effects in general relativity such as the Hawking effect (see Hawking (1975), Bachelot (1997), (1999), (2000), and Melnyk (2004)). Finally, the third motivation is dictated by the study of resonances, i.e. of the complex frequencies which are poles of the analytic continuation of the scattering operator. For such studies in the Schwarzschild geometry we refer to Bachelot, and Motet-Bachelot (1993), and S$\grave{\mbox{a}}$ Barreto, and Zworski (1998).\\
Concerning the time-dependent scattering theory for Dirac particles in a Coulomb field we find some early publications by Dollard (1964), by Dollard, and Velo (1966), and by Enss, and Thaller (1986). The time-dependent scattering theory for classical, and quantum scalar fields on the Schwarzschild metric was first obtained by Dimock (1985), and by Dimock, and Kay (1986a,b), (1987). In the same geometry Bachelot developed the scattering theory for electromagnetic fields (1991), and Klein-Gordon fields (1994). Regarding Dirac fields in the Schwarzschild geometry, Nicolas (1995a) presented a scattering theory for classical massless Dirac particles, and Jin (1998) constructed wave operators, classical at the event horizon and Dollard-modified at infinity, in the massive case.  Moreover, Melnyk (2003) gave a complete scattering theory for massive charged Dirac fields in the Reissner-Nordstr$\phi$m metric. Finally, Daud$\acute{\mbox{e}}$ (2004) proved the existence and asymptotic completeness of wave operators, classical at the event horizon, and Dollard-modified at infinity, for classical massive Dirac particles in the Kerr-Newman geometry by means of the Mourre theory (see Mourre (1981)). For the nonlinear Klein-Gordon equation on Schwarzschild like metrics partial scattering results by means of conformal methods have been obtained by Nicolas (1995b). A complete scattering theory for the wave equation, on stationary, asymptotically flat space-times, was developed by H\"{a}fner (2001).\\
Whenever we attempt to analyze the scattering properties of fields outside the event horizon of a Kerr black hole, we are faced with some difficulties which are not present in the picture of the Schwarzschild metric. First of all, the Kerr solution is only axially symmetric since it possesses only two commuting Killing vector fields, namely the time coordinate vector field $\partial_{t}$ and the longitude coordinate vector field $\partial_{\varphi}$. This implies that there is no decomposition in spin-weighted spherical harmonics. Moreover, another apparent difficulty is due to fact that it is impossible to find a Killing vector field which is time-like everywhere outside the black hole. In fact $\partial_{t}$ becomes space-like in the ergo-sphere, a toroidal region around the horizon. This implies that for field equations describing particles of integer spin (wave equation, Klein-Gordon, Maxwell) there exists no positive definite conserved energy. For field equations describing particles with half-integer spin (Weyl, Dirac) we can find a conserved $L_{2}$ norm with the usual interpretation of a conserved charge. Hence, the absence of stationarity in the Kerr metric is not really a difficulty for the scattering theory of classical Dirac fields. Nevertheless there are only few analytical studies of the propagation of fields outside Kerr black holes.\\
Our work represents a new approach to the results obtained by Jin (1998), and by H\"{a}fner, and Nicolas (2004). Firstly, our method is based on an integral representation for the Dirac propagator outside the event horizon of a non-extreme Kerr manifold: this is new in this context. Secondly, we are able to compute explicitly the wave operators (Dollard-modified) at infinity. Moreover, by computing the wave operators (classical) at the event horizon and introducing suitable global wave operators, it should be possible to give an alternative proof for the asymptotic completeness to that one presented by H\"{a}fner, and Nicolas, and to calculate the scattering matrix. This will be done in the next future. \\
Let us briefly describe the contents of this paper. In Section~\ref{sec:2} we give the integral representation for the Dirac propagator in the exterior region of a Kerr manifold, and the asymptotic behavior for the radial solutions. These results are essential to the development of our theory. In Section~\ref{sec:3} we define the free dynamics asymptotically at infinity, and we introduce the so-called Dollard-modified wave operators. In Section~\ref{sec:4} we compute explicitly the phase shift we need to implement in the free dynamics. In this section the main result is Theorem~\ref{THMI} where we give an integral representation for the Dollard-modified wave operators.    

\section{\label{sec:2}Preliminaries}
In Boyer-Lindquist coordinates $(t,r,\vartheta,\varphi)$ with $r>0$, $0\leq\vartheta\leq\pi$, $0\leq\varphi<2\pi$ the Kerr metric is given by (e.g. Wald)
\begin{equation}\label{KN}
ds^{2}=\frac{\Delta-a^{2}\sin^{2}{\vartheta}}{\Sigma}dt^{2}+\frac{2a\sin^{2}{\vartheta}(r^2+a^2-\Delta)}{\Sigma}dtd\varphi-\frac{\Sigma}{\Delta}dr^2-\Sigma d\vartheta^{2}-(r^2+a^2)^{2}\frac{\widetilde{\Sigma}}{\Sigma}d\varphi^{2}
\end{equation}
with
\[
\Sigma:=\Sigma(r,\theta)=r^2+a^2\cos^{2}\theta, \qquad \Delta:=\Delta(r)=r^2-2Mr+a^2,
\]
and
\[
\widetilde{\Sigma}:=\widetilde{\Sigma}(r,\vartheta)=1-a^2\gamma^{2}(r)\sin^{2}{\vartheta},\qquad \gamma(r):=\frac{\sqrt{\Delta}}{r^2+a^2}
\]
where $M$, and $a$ are the mass, and the angular momentum per unit mass of the black hole, respectively. Here, $a$ is allowed to be zero, so that our results apply also to the Schwarzschild metric. Moreover, we will always work in the non-extreme case $M^2>a^2$ which implies that the function $\Delta$ has two distinct zeros at the Cauchy horizon $r_{0}=M-\sqrt{M^2-a^2}$, and at the event horizon $r_{1}=M+\sqrt{M^2-a^2}$. Notice that $\Delta>0$ for $r>r_{1}$. Outside the event horizon of a non-extreme Kerr manifold the integral representation of the Dirac propagator for a particle of mass $m_{e}$, charge $e$, and energy $\omega$ is (Theorem 5.4 in Batic, and Schmid (2006))
\begin{equation}\label{rappresentazione}
\psi(t,x)=\int_{-\infty}^{+\infty}d\omega\,e^{i\omega t}\sum_{j\in\mathbb{Z}\backslash\{0\}}\sum_{k\in\mathbb{Z}}\psi_{\omega}^{kj}(x)\langle\psi^{kj}_{\omega}|\psi_{0}\rangle,\quad x=(u,\vartheta,\varphi)
\end{equation}
where $\langle\cdot|\cdot\rangle$ is the positive definite scalar product
\begin{equation}\label{scalar1}
\langle\psi^{kj}_{\omega}|\psi_{0}\rangle=\int_{-\infty}^{+\infty}\,du\int_{-1}^{+1}\,d(\cos{\vartheta})\int_{0}^{2\pi}\,d\varphi \hspace{1mm}\sqrt{\widetilde{\Sigma}}\hspace{2mm}\overline{\psi}^{\hspace{1mm}kj}_{\omega}\psi_{0},
\end{equation}
and $\psi_{0}$ denotes some initial data in $\mathcal{C}^{\infty}_{0}(\Omega)^{4}$ with $\Omega:=\mathbb{R}\times S^{2}$. Here, $u\in\mathbb{R}$ is the so-called tortoise coordinate defined by $du/dr=(r^{2}+a^{2})/\Delta$. Notice that the variable $u$ has the property to approach $-\infty$ as $r$ approaches the event horizon. Furthermore, $k\in\mathbb{Z}$ is the azimuthal quantum number, $j\in\mathbb{Z}\backslash\{0\}$ labels the eigenvalues of the angular operator arising from the Chandrasekhar separation of the Dirac equation into a radial, and an angular system of ODEs (see Chandrasekhar (1976)), and according to the Chandrasekhar ansatz $\psi_{\omega}^{kj}$ has the following form
\begin{equation}\label{Ansatz}
\psi^{kj}_{\omega}(x)=\frac{1}{\sqrt{2\pi}}\left( \begin{array}{c}
               R_{\omega,-}^{kj}(u)S_{\omega,-}^{kj}(\vartheta)\\
               R_{\omega,+}^{kj}(u)S_{\omega,+}^{kj}(\vartheta)\\
               R_{\omega,+}^{kj}(u)S_{\omega,-}^{kj}(\vartheta)\\
               R_{\omega,-}^{kj}(u)S_{\omega,+}^{kj}(\vartheta)
\end{array} \right)e^{i\left(k+\frac{1}{2}\right)\varphi}
\end{equation}
with $R^{kj}_{\omega}=(R_{\omega,-}^{kj},R_{\omega,+}^{kj})^{T}$, and $S^{kj}_{\omega}=(S_{\omega,-}^{kj},S_{\omega,+}^{kj})^{T}$ the radial, and angular components of the spinor $\psi_{\omega}^{kj}$, respectively. For further properties on the angular eigenfunctions $S^{kj}_{\omega}$ we refer to Finster et al. (2000), and to Batic et al. (2005). As shown in Lemma 6.1-2 in Batic and, Schmid (2006) in the limit $u\to-\infty$ the radial solutions behave for $u\leq u_{1}<0$ as follows
\begin{equation}\label{radialmin}
R^{kj}_{\omega}(u)= \left( \begin{array}{cc}
                                     e^{-i\Omega_{0}u}[f^{(0)}_{-}+\mathcal{O}(e^{du})] \\
                                     e^{+i\Omega_{0}u}[f^{(0)}_{+}+\mathcal{O}(e^{du})]
                                     \end{array}\right)
\end{equation}
with $f^{(0)}_{\pm}=c_{\pm}^{(0)}e^{\mp i\Omega_{0}u_{1}}$ such that $|c^{(0)}_{-}|^2+|c^{(0)}_{+}|^2\neq 0$, and
\[
\Omega_{0}=\omega+\frac{a\left(k+\frac{1}{2}\right)}{r_{1}^{2}+a^{2}},\quad 0<d=4\kappa_{+},\quad\kappa_{+}=\frac{r_{1}-r_{0}}{2(r_{1}^2+a^2)}
\] 
where $\kappa_{+}$ is the surface gravity at $r=r_{1}$. For $u\to+\infty$ we have to distinguish between the cases $|\omega|>m_{e}$, and $|\omega|<m_{e}$. In the first case the asymptotic behavior of the radial functions is for $u\geq u_{0}>0$
\begin{equation}\label{radialmag}
R^{kj}_{>}(u)=\mathcal{T}(\omega)\left( \begin{array}{cc}
                                     e^{-i\Phi(u)}[f^{\infty}_{-}+\frac{D_{-,kj}^{(1)}(\omega)}{u}+\mathfrak{R}^{kj}_{-,\omega}(u)] \\
                                     e^{+i\Phi(u)}[f^{\infty}_{+}+\frac{D_{+,kj}^{(1)}(\omega)}{u}+\mathfrak{R}^{kj}_{+,\omega}(u)]
                                     \end{array}\right),\quad|\mathfrak{R}^{kj}_{\pm,\omega}(u)|\leq\frac{|D_{\pm,kj}^{(2)}(\omega)|}{u^2}
\end{equation}
where $f^{\infty}_{\pm}=c_{\pm}e^{\mp i\Phi(u_{0})}$ with $c_{\pm}\in\mathbb{C}$ such that $|c_{-}|^2+|c_{+}|^2\neq 0$,
\begin{eqnarray*}
\mathcal{T}(\omega)&=&\left( \begin{array}{cc}
     \cosh{\Theta}&\sinh{\Theta}\\
     \sinh{\Theta}&\cosh{\Theta}
           \end{array} \right),\quad\Theta=\frac{1}{4}\log{\left(\frac{\omega+m_{e}}{\omega-m_{e}}\right)},\quad\Phi(u)=\kappa u+\frac{Mm^{2}_{e}}{\kappa}\log{u},\\
\kappa&=&\epsilon(\omega)\sqrt{\omega^2-m^2_{e}},\quad\epsilon(\omega)=\left\{\begin{array}{ll}
            +1&\mbox{if $\omega>+m_{e}$}\\
            -1&\mbox{if $\omega<-m_{e}$}
            \end{array}\right.,
\end{eqnarray*}
and
\[
D_{\pm,kj}^{(1)}(\omega)=\pm i\widetilde{P}_{kj}(\omega),\quad D_{\pm,kj}^{(2)}(\omega)=-\frac{\widetilde{P}^{2}_{kj}(\omega)}{2}\mp iP_{kj}(\omega)
\]
with
\begin{eqnarray*}
\widetilde{P}_{kj}(\omega)&=&\frac{\kappa}{2}\left(\frac{\widehat{P}_{kj}(\omega)}{\kappa^{2}}+\frac{M^{2}m^{4}_{e}}{\kappa^{4}}\right),\quad \widehat{P}_{kj}(\omega)=\lambda^{2}_{j}(\omega)-m^{2}_{e}a^{2}-2\left(k+\frac{1}{2}\right)a\omega\\
P_{kj}(\omega)&=&\frac{\kappa}{4}\left(\frac{2M(\lambda^{2}_{j}(\omega)-2m^{2}_{e}a^{2})}{\kappa^{2}}+\frac{Mm^{2}_{e}\widehat{P}_{kj}(\omega)}{\kappa^{4}}+\frac{M^{3}m_{e}^{6}}{\kappa^{6}}\right).
\end{eqnarray*}
Notice that for some $\epsilon>0$ $R^{kj}_{>}(u)$ is smooth in $\omega\in(-\infty,-m_{e}-\epsilon]\cup[m_{e}+\epsilon,+\infty)$. For $|\omega|<m_{e}$ there are two fundamental solutions with exponential decay, and growth, respectively which are smooth in $\omega\in[-m_{e}+\epsilon,m_{e}-\epsilon]$. In order to disregard the unphysical solution with exponential growth we normalize the radial solutions $R^{kj}_{<}(u)$ by imposing that $\left|R^{kj}_{<}(u)\right|\to 1$ as $u\to\infty$.

\section{\label{sec:3}Dollard-modified wave operators}
Because of the presence of two asymptotic regions ($u\to\pm\infty$) we need to specify for each of them an asymptotic dynamics. For $u\to+\infty$ we define the free dynamics by replacing the Hamiltonian (4.2) in Batic, and Schmid (2006) by its formal limit $H_{\infty}$ when $M\to 0$. Notice that in this case the Kerr metric goes over to the Minkowski metric in oblate spheroidal coordinates (OSC). Since $du/dr\hspace{1mm}=1$ for $M=0$, and $r$ can be extended to negative values, we can identify the tortoise coordinate $u$ with the spatial variable $r$. We consider $H_{\infty}$ as an operator acting on the Hilbert space
\[
\mathcal{H}_{\infty}=L_{2}\left(\Omega,d\mu_{\infty}\right)^{4},\quad d\mu_{\infty}=\sqrt{\frac{\Sigma}{u^2+a^2}}\hspace{1mm}du\hspace{1mm}d(\cos{\vartheta})\hspace{1mm}d\varphi.
\]
Moreover, the Hamiltonian $H_{\infty}$ is formally self-adjoint with respect to the positive scalar product (Sec. 7 ibid.)
\begin{equation}\label{scalar1OSC}
\langle\psi^{(\infty)}|\phi^{(\infty)}\rangle_{(\infty)}=\int_{\Omega}\,d\mu_{\infty}\hspace{1mm}\overline{\psi}^{\hspace{1mm}(\infty)}\phi^{(\infty)},\quad \psi^{(\infty)},\phi^{(\infty)}\in\mathcal{H}_{\infty},
\end{equation}
it is essentially self-adjoint on $\mathcal{C}_{0}^{\infty}(\Omega,d\mu_{\infty})^{4}$, and it has a unique self-adjoint extension on the Sobolev space $W^{1,2}(\Omega,d\mu_{\infty})^{4}$.  The Dirac propagator for a particle of mass $m_{e}$, charge $e$, and energy $\omega$ in the Minkowski metric expressed in OSC is (Lemma 7.2 ibid.)
\begin{equation}\label{rappresentazioneOSC}
\psi^{(\infty)}(t,x)=\int\limits_{\sigma(H_{\infty})}d\omega\,e^{i\omega t}\sum_{j\in\mathbb{Z}\backslash\{0\}}\sum_{k\in\mathbb{Z}}\psi_{\omega}^{kj,(\infty)}(x)\langle\psi^{kj,(\infty)}_{\omega}|\psi_{0}^{(\infty)}\rangle,\quad x=(u,\vartheta,\varphi)
\end{equation}
where $\psi_{0}^{(\infty)}\in\mathcal{C}^{\infty}_{0}(\Omega,d\mu_{\infty})^{4}$, $\sigma(H_{\infty})=(-\infty,-m_{e}]\cup[m_{e},+\infty)$, and 
\begin{equation}\label{AnsatzOSC}
\psi^{kj,(\infty)}_{\omega}(x)=\frac{1}{\sqrt{2\pi}}\left( \begin{array}{c}
               R_{\omega,-}^{kj,(\infty)}(u)S_{\omega,-}^{kj}(\vartheta)\\
               R_{\omega,+}^{kj,(\infty)}(u)S_{\omega,+}^{kj}(\vartheta)\\
               R_{\omega,+}^{kj,(\infty)}(u)S_{\omega,-}^{kj}(\vartheta)\\
               R_{\omega,-}^{kj,(\infty)}(u)S_{\omega,+}^{kj}(\vartheta)
\end{array} \right)e^{i\left(k+\frac{1}{2}\right)\varphi}.
\end{equation}
Notice that the angular components of the spinor $\psi_{\omega}^{kj,(\infty)}$ are the same as those for $\psi_{\omega}^{kj}$ since the angular operator arising from the Chandrasekhar separation of the Dirac equation into a radial, and an angular system of ODEs does not contain the mass parameter M. The asymptotic behavior of the radial functions is (Lemma 7.3 ibid.)
\begin{equation}\label{radialOSC}
R^{kj,(\infty)}_{\omega}(u)=\mathcal{T}(\omega)\left( \begin{array}{cc}
                                     e^{-i\kappa u}[\widetilde{f}^{(\infty)}_{-}+\frac{D_{-,kj}^{(1),(\infty)}(\omega)}{u}+\mathfrak{R}^{kj,(\infty)}_{-,\omega}(u)] \\
                                     e^{+i\kappa u}[\widetilde{f}^{(\infty)}_{+}+\frac{D_{+,kj}^{(1),(\infty)}(\omega)}{u}+\mathfrak{R}^{kj,(\infty)}_{+,\omega}(u)]
                                     \end{array}\right),\quad|\mathfrak{R}^{kj,(\infty)}_{\pm,\omega}(u)|\leq\frac{|D_{\pm,kj}^{(2),(\infty)}(\omega)|}{u^2}
\end{equation}
where $\widetilde{f}^{(\infty)}_{\pm}=\widetilde{c}_{\pm}e^{\mp i\kappa u_{0}}$ with $\widetilde{c}_{-}$, $\widetilde{c}_{+}\in\mathbb{C}$ such that $|\widetilde{c}_{-}|^2+|\widetilde{c}_{+}|^2\neq 0$. Moreover, $\Theta$, and $\kappa$ are defined as in Section~\ref{sec:2}, and
\[
D_{\pm,kj}^{(1),(\infty)}(\omega)=\pm i\frac{\widehat{P}_{kj}(\omega)}{2\kappa},\quad D_{\pm,kj}^{(2),(\infty)}(\omega)=-\frac{\widehat{P}^{2}_{kj}(\omega)}{8\kappa^{2}}.
\]
Clearly, $R^{kj,(\infty)}_{\omega}(u)$ is smooth in $\omega\in(-\infty,-m_{e}-\widetilde{\mu}]\cup[m_{e}+\widetilde{\mu},+\infty)$ with $\widetilde{\mu}>0$. In what follows we consider the full Hamiltonian $H$ as an operator acting on the Hilbert space $\mathcal{H}=L_{2}(\Omega,d\mu)^{4}$ with $d\mu=\sqrt{\widetilde{\Sigma}}\hspace{1mm}du\hspace{1mm}d(\cos{\vartheta})\hspace{1mm}d\varphi$. Since we want to compare $H$ with $H_{\infty}$ we define the smooth bounded identifying operator $\mathcal{I}_{\infty}:\mathcal{H}_{\infty}\longrightarrow\mathcal{H}$ as follows
\begin{equation}\label{IINFHOR}
\mathcal{I_{\infty}}\psi^{(\infty)}=\chi_{\infty}(u)\psi^{(\infty)},\quad\psi^{(\infty)}\in\mathcal{H}_{\infty}
\end{equation}
with cut-off function $\chi_{\infty}\in\mathcal{C}^{\infty}(\mathbb{R})$ such that
\begin{equation}\label{cutoff_0}
\chi_{\infty}(u)=\left\{\begin{array}{lll}
                                        
                                        1 & \mbox{if $u\geq\widehat{u}_{0}$} \\
                                        \eta_{\infty}(u) & \mbox{if $u\in[u_{0},\widehat{u}_{0}]$}\\
                                        0 & \mbox{if $u\leq u_{0}$}                  
                   \end{array}\right. 
\end{equation}
with $0\leq\eta_{\infty}(u)\leq 1$ for $\widehat{u}_{0}\geq u\geq u_{0}>0$. Because of the long-range nature of the gravitational potential we define asymptotically at infinity the following Dollard-modified wave operators (see Dollard (1964))
\begin{equation} \label{modificato}
W^{\pm}_{(\infty)}\psi_{0}^{(\infty)}=s-\lim_{t\to \pm \infty}e^{-iHt}\mathcal{I}_{\infty} e^{iH_{\infty}t}e^{i\delta(t)}\psi_{0}^{(\infty)},\quad \psi_{0}^{(\infty)}\in\mathcal{H}_{\infty}
\end{equation}
under the conditions that the phase shift operator $\delta(t)$ commutes with $H_{\infty}$, and that the limit exists in the strong sense in $\mathcal{H}$. If we apply \eqref{rappresentazione}, and \eqref{rappresentazioneOSC} we can express \eqref{modificato} as follows 
\begin{multline}\label{INTAI}
\left(W^{\pm}_{(\infty)}\psi_{0}^{(\infty)}\right)(x^{'})=\lim_{t\to \pm\infty}\int\limits_{\sigma(H)}\,d\omega^{'}\sum_{j^{'}\in\mathbb{Z}\backslash\{0\}}\sum_{k^{'}\in\mathbb{Z}}\psi^{k^{'}j^{'}}_{\omega^{'}}(x^{'})\int\limits_{\Omega}\,d\mu\chi_{\infty}(u)\int\limits_{\sigma(H_{\infty})}\,d\omega\hspace{1mm}e^{i(\omega-\omega^{'}) t+i\delta(t)}\\
\sum_{j\in\mathbb{Z}\backslash\{0\}}\sum_{k\in\mathbb{Z}}\overline{\psi}^{\hspace{1mm}k^{'}j^{'}}_{\omega^{'}}(x)\hspace{1mm}\psi^{kj,(\infty)}_{\omega}(x)\langle\psi^{kj,(\infty)}_{\omega}|\psi^{(\infty)}_{0}\rangle_{(\infty)}.
\end{multline}
In preparation of the next results we define the intervals $B_{\epsilon}(\pm m_{e}):=\{\omega^{'}\in\mathbb{R}||\omega^{'}\pm m_{e}|<\epsilon\}$, $I_{+}:=\{\omega\in\sigma(H_{\infty})|0\leq\omega-m_{e}<\widetilde{\mu}\}$, and $I_{-}:=\{\omega\in\sigma(H_{\infty})|-\widetilde{\mu}\leq\omega+m_{e}\leq 0\}$ for some $\epsilon>0$. The following lemma controls uniformly in $t$ the contributions of the frequency sets $B_{\epsilon}(\pm m_{e})$, and $I_{\pm}$ to the wave operators asymptotically at infinity. 
\begin{lemma} \label{pmm}
Let $E:\mathbb{R}\longrightarrow\mathcal{B}(\mathcal{H})$, and $E^{\infty}:\sigma(H_{\infty})\longrightarrow\mathcal{B}(\mathcal{H}_{\infty})$ be the spectral families associated to the self-adjoint operators $H$ and $H_{\infty}$, respectively. In particular, let us consider their restrictions on the Borel sets $B_{\epsilon}:=B_{\epsilon}(-m_{e})\cup B_{\epsilon}(m_{e})$, and $I:=I_{+}\cup I_{-}$, i.e. $E_{\epsilon}:B_{\epsilon}\longrightarrow\mathcal{B}(\mathcal{H})$, $E_{\widetilde{\mu}}^{\infty}:I\longrightarrow\mathcal{B}(\mathcal{H}_{\infty})$ with
\[
E_{\epsilon}:=\int\limits_{\mathbb{R}}\,\chi_{B_{\epsilon}}(\omega^{'})\hspace{1mm}dE(\omega^{'}),\quad E_{\widetilde{\mu}}^{\infty}:=\int\limits_{\sigma(H_{\infty})}\,\chi_{I}(\omega)\hspace{1mm}dE^{\infty}(\omega)
\]
where $\chi_{B_{\epsilon}}$, and $\chi_{I}$ are the characteristic functions of the intervals $B_{\epsilon}$, and $I$. Then, for every $\kappa>0$, and $\psi_{0}^{(\infty)}\in\mathcal{C}_{0}^{\infty}(\Omega,d\mu_{\infty})^{4}$ there exist constants $\widetilde{\mu}$, $\epsilon>0$ such that for every $t$
\[
\|U_{\infty}(t)\psi_{0}^{(\infty)}-e^{-iHt}(Id-E_{\epsilon})\mathcal{I}_{\infty}e^{iH_{\infty}t}(Id_{\infty}-E_{\widetilde{\mu}}^{\infty})e^{i\delta(t)}\psi_{0}^{(\infty)}\|_{\mathcal{H}}<\kappa
\]
where $U_{\infty}(t):=e^{-iHt}\mathcal{I}_{\infty}e^{iH_{\infty}t}e^{i\delta(t)}$, and $\delta(t)$ is a phase shift operator commuting with $H_{\infty}$.
\end{lemma}
\begin{proof}
Since there is no risk of confusion we omit to write explicitly the superscript $(\infty)$ attached to the initial data. Let us define
\[
\Delta W:=e^{-iHt}\mathcal{I}_{\infty}e^{iH_{\infty}t}e^{i\delta(t)}\psi_{0}-e^{-iHt}(Id-E_{\epsilon})\mathcal{I}_{\infty}e^{iH_{\infty}t}(Id_{\infty}-E_{\widetilde{\mu}}^{\infty})e^{i\delta(t)}\psi_{0}.
\]
By adding and subtracting the term $e^{-iHt}\mathcal{I}_{\infty}e^{iH_{\infty}t}(Id_{\infty}-E_{\widetilde{\mu}}^{\infty})e^{i\delta(t)}\psi_{0}$ to $\Delta W$ we obtain
\[
\Delta W=e^{-iHt}\mathcal{I}_{\infty}e^{iH_{\infty}t}e^{i\delta(t)}E^{\infty}_{\widetilde{\mu}}\psi_{0}+e^{-iHt}E_{\epsilon}\mathcal{I}_{\infty}e^{iH_{\infty}t}e^{i\delta(t)}\psi_{0}-e^{-iHt}E_{\epsilon}\mathcal{I}_{\infty}e^{iH_{\infty}t}E^{\infty}_{\widetilde{\mu}}e^{i\delta(t)}\psi_{0}.
\]
 Since $e^{-iHt}$, and $e^{iH_{\infty}t}$ are unitary, $\mathcal{I}_{\infty}$ is bounded, $e^{i\delta(t)}$ commutes with $E_{\widetilde{\mu}}^{\infty}$, and $E_{\epsilon}$ commutes with $e^{-iHt}$, we have
\begin{equation} \label{Gleich1}
\|\Delta W\|_{\mathcal{H}}\leq 2\|E_{\widetilde{\mu}}^{\infty}\psi_{0}\|_{\mathcal{H}_{\infty}}+\|E_{\epsilon}U_{\infty}(t)\psi_{0}\|_{\mathcal{H}}.
\end{equation}
Let us recall that
\[
E_{\widetilde{\mu}}^{\infty}\psi_{0}=\sum_{\pm}\int\limits_{\sigma(H_{\infty})}\,d\omega\chi_{I_{\pm}}(\omega)~dE^{(\infty)}(\omega)\psi_{0}.
\]
Without loss of generality we consider the above expression for the set $I_{+}$. Taking into account that the idempotent projector $E_{\widetilde{\mu}}^{+,\infty}$ is hermitian, and making use of (5.8) in Batic, and Schmid (2006), it follows that 
\begin{multline*}
\|E_{\widetilde{\mu}}^{+,\infty}\psi_{0}\|_{\mathcal{H}_{\infty}}^{2}=\langle E_{\widetilde{\mu}}^{+,\infty}\psi_{0}|E_{\widetilde{\mu}}^{+,\infty}\psi_{0}\rangle_{\mathcal{H}_{\infty}}=\langle \psi_{0}|E_{\widetilde{\mu}}^{+,\infty}\psi_{0}\rangle_{\mathcal{H}_{\infty}}=\int\limits_{\sigma(H_{\infty})}\,\chi_{I_{+}}(\omega)~d\langle \psi_{0}|E^{\infty}(\omega)\psi_{0}\rangle\\
=\int\limits_{\sigma(H_{\infty})}\,d\omega\chi_{I_{+}}(\omega)\frac{d}{d\omega}\langle \psi_{0}|E^{\infty}(\omega)\psi_{0}\rangle=\int\limits_{\sigma(H_{\infty})}\,\hspace{1mm}d\omega\chi_{I_{+}}(\omega)\langle\psi_{0,\omega}|\psi_{0,\omega}\rangle_{\mathfrak{h}_{\infty}}
\end{multline*}
where $\psi_{0,\omega}=(\mathcal{F}\psi_{0})(\omega)\in\mathfrak{h}_{\infty}$ is the representative of the element $\psi_{0}\in\mathcal{H}_{\infty}$, the map $\mathcal{F}:\mathcal{H}_{\infty}\longrightarrow\mathfrak{H}_{\infty}$ is unitary, and $\langle\psi_{0,\omega}|\psi_{0,\omega}\rangle_{\mathfrak{h}_{\infty}}=|\psi_{0,\omega}|^{2}$. For the definition of $\mathfrak{h}_{\infty}$ see Theorem 5.2 (ibid.). Thus, we obtain
\[
\|E_{\widetilde{\mu}}^{\infty}\psi_{0}\|_{\mathcal{H}_{\infty}}^{2}=\int\limits_{\sigma(H_{\infty})}\,d\omega\chi_{I}(\omega)|\psi_{0,\omega}|^{2}.
\]
Since the r.h.s. of the above expression converges to zero with the length of $I$, by choosing $\widetilde{\mu}$ sufficiently small we can arrange that
\begin{equation}\label{stima1}
\|E_{\widetilde{\mu}}^{\infty}\psi_{0}\|_{\mathcal{H}_{\infty}}<\kappa/4.
\end{equation}
According to Daud$\acute{\mbox{e}}$ (2004) the strong limit of $U_{\infty}$ exists. Moreover, Lemma 5.3 in Batic and Schmid (2006) implies that the point spectrum of $H$ is empty. Hence, for a fixed $\phi\in\mathcal{H}$ it results that $\|E_{\epsilon}\phi\|_{\mathcal{H}}\to 0$ for $\epsilon\to 0$. Such a limit is uniform for $\phi\in C\subset\mathcal{H}$ with $C$ compact. Since $U_{\infty}(t)\psi_{0}$ is continuous in $t$ and converges for $t\to\pm\infty$, the closure of $\{U_{\infty}(t)\psi_{0}\}$ is a compact set. Thus, for a given $k>0$ we can always choose $\epsilon$ sufficiently small such that for every $t\in\mathbb{R}$
\begin{equation}\label{stima2}
\|E_{\epsilon}U_{\infty}(t)\psi_{0}\|_{\mathcal{H}}<\frac{k}{2}.
\end{equation}
The estimates \eqref{stima1}, and \eqref{stima2} together with \eqref{Gleich1} complete the proof.\hspace{5mm}$\square$
\end{proof}
Let us now introduce the set $\Lambda=\{\omega^{'}\in\mathbb{R}|-\omega_{0}\leq\omega^{'}\leq\omega_{0}\}$ for some $\omega_{0}>m_{e}$. In the following remark we explain how to bring the limit $t\to\pm\infty$ inside the integral over $\omega^{'}$, and why we can consider a finite number of terms in the sums over $j^{'}$, $k^{'}$, $j$, $k$ entering in \eqref{INTAI}.
\begin{remark}\label{xc}
Let $E$ be the spectral family defined in Lemma~\ref{pmm}, and let us consider its restriction on the set $\Lambda$
\[
E_{\Lambda}:\Lambda\longrightarrow\mathcal{B}(\mathcal{H}),\quad E_{\Lambda}:=\int\limits_{\mathbb{R}}\,\chi_{\Lambda}(\omega^{'})\hspace{1mm}dE(\omega^{'})
\]
where $\chi_{\Lambda}$ is the characteristic function of the interval $\Lambda$. Since $U_{\infty}(t)$ converges strongly to $W^{\pm}_{(\infty)}$ for $t\to\pm\infty$ (see Daud$\acute{\mbox{e}}$ (2004)), and $E_{\Lambda}U_{\infty}(t)$ converges to $E_{\Lambda}W^{\pm}_{(\infty)}$ for $t\to\pm\infty$ it follows that
\[
E_{\Lambda}W^{\pm}_{(\infty)}=s-\lim_{t\to \pm \infty}E_{\Lambda}e^{-iHt}\mathcal{I}_{\infty} e^{iH_{\infty}t}e^{i\delta(t)}.
\]
Finally, taking into account that the spectrum of $H$ is purely absolutely continuous, and that $E_{\Lambda}$ converges strongly to the identity for $\omega_{0}\to+\infty$ it results that $E_{\Lambda}W^{\pm}_{(\infty)}$ converges strongly to $W^{\pm}_{(\infty)}$ for $\omega_{0}\to+\infty$. By means of a similar argument it can be shown that we can restrict our attention to a finite number of quantum numbers $j^{'}$, $k^{'}$, $j$, $k$.
\end{remark}

\section{\label{sec:4} The wave operators asymptotically at infinity}
Let us define the sets $\Omega_{I}:=[-\omega_{0},-m_{e}-\epsilon]\cup [m_{e}+\epsilon,\omega_{0}]$, $\Omega_{II}:=[-m_{e}+\epsilon,m_{e}-\epsilon]$, and $\Omega_{III}:=(-\infty,-m_{e}-\widetilde{\mu}]\cup[m_{e}+\widetilde{\mu},+\infty)$. Taking into account that $\sqrt{\widetilde{\Sigma}}$ admits the following asymptotic expansion in the spatial variable $u$
\[
\sqrt{\widetilde{\Sigma}}=1+\mathcal{R}(u,\vartheta),\quad |\mathcal{R}(u,\vartheta)|\leq\frac{a^{2}}{u^{2}}
\]
if we apply Lemma~\ref{pmm}, Remark~\ref{xc}, and if we make use of \eqref {Ansatz}, \eqref{radialmag}, \eqref{radialOSC}, and Lemma 6.1 in Batic, and Schmid (2006), concerning the asymptotic behavior of the radial functions $R^{kj}_{<}(u)$, then we end up after a lengthy computation with the following expression for the Dollard-modified wave operator \eqref{INTAI}  
\[
W^{\pm}_{(\infty)}\psi^{(\infty)}_{0}=W^{\pm}_{(\infty),I}\psi^{(\infty)}_{0}+W^{\pm}_{(\infty),II}\psi^{(\infty)}_{0}
\]
with
\begin{multline}\label{madre}
W^{\pm}_{(\infty),I/II}\psi^{(\infty)}_{0}=\int\limits_{\Omega_{I/II}}\,d\omega^{'}\sum_{|j^{'}|\leq j^{'}_{0}}\sum_{|j|\leq j_{0}}\sum_{|k|\leq k_{0}}\psi^{kj^{'}}_{\omega^{'}}(x)\lim_{t\to \pm\infty}\int\limits_{\mathbb{R}}\,du\chi_{\infty}(u)\\
\int\limits_{\Omega_{III}}\,d\omega\hspace{1mm}e^{i(\omega-\omega^{'}) t+i\delta(t)}F^{kjj^{'}}_{I/II}(\omega^{'},\omega,u),
\end{multline}
\begin{multline} \label{FI}
F^{kjj^{'}}_{I}(\omega^{'},\omega,u)=\sum\limits_{\pm}\left\{e^{\pm i(\Phi(u)-\kappa u)}Z(\omega^{'},\omega)\right.\left.\left[A_{kjj^{'}}\left(\overline{f}^{\infty}_{\mp}\widetilde{f}^{\infty}_{\mp}+\frac{\mathfrak{A}^{kjj^{'}}_{\pm}}{u}\right)+h_{\pm}^{kjj^{'}}(\omega^{'},\omega,u)\right]\right.\\
\left. +e^{\pm i(\Phi(u)+\kappa u)}X(\omega^{'},\omega)\left[A_{kjj^{'}}\left(\overline{f}^{\infty}_{\mp}\widetilde{f}^{\infty}_{\pm}+\frac{\mathfrak{B}^{kjj^{'}}_{\pm}}{u}\right)+\widehat{h}_{\pm}^{kjj^{'}}(\omega^{'},\omega,u)\right]\right\}g^{kj}(\omega),
\end{multline}
\begin{multline} \label{FII}
F^{kjj^{'}}_{II}(\omega^{'},\omega,u)=e^{-\widehat{\Phi}(u)}\left\{e^{-i\kappa u}\mathcal{Z}(\omega^{'},\omega)\right.\left.\left[A_{kjj^{'}}\left(\overline{\widehat{f}}^{\hspace{1mm}\infty}_{-}\widetilde{f}^{\infty}_{-}+\frac{\widehat{\mathfrak{A}}^{kjj^{'}}_{\pm}}{u}\right)+\tau_{\pm}^{kjj^{'}}(\omega^{'},\omega,u)\right]+\right.\\
\left. +e^{i\kappa u}\mathcal{X}(\omega^{'},\omega)\left[A_{kjj^{'}}\left(\overline{\widehat{f}}^{\hspace{1mm}\infty}_{-}\widetilde{f}^{\infty}_{+}+\frac{\widehat{\mathfrak{B}}^{kjj^{'}}_{\pm}}{u}\right)+\widehat{\tau}_{\pm}^{kjj^{'}}(\omega^{'},\omega,u)\right]\right\}g^{kj}(\omega)
\end{multline}
where $g^{kj}(\omega)=\langle\psi^{kj,(\infty)}_{\omega}|\psi^{(\infty)}_{0}\rangle_{(\infty)}$, and
\begin{eqnarray*}
Z(\omega^{'},\omega)&=&\cosh{\Theta(\omega^{'})}\cosh{\Theta(\omega)}+\sinh{\Theta(\omega^{'})}\sinh{\Theta(\omega)},\\
X(\omega^{'},\omega)&=&\cosh{\Theta(\omega^{'})}\sinh{\Theta(\omega)}+\sinh{\Theta(\omega^{'})}\cosh{\Theta(\omega)},\\
\mathcal{Z}(\omega^{'},\omega)&=&\sinh{\Theta(\omega)}+\tau(\omega^{'})\cosh{\Theta(\omega)},\quad\tau(\omega^{'})=\frac{1}{m_{e}}\left(\omega^{'}+i\sqrt{m^{2}_{e}-{\omega^{'}}^{2}}\right)\\
\mathcal{X}(\omega^{'},\omega)&=&\cosh{\Theta(\omega)}+\tau(\omega^{'})\sinh{\Theta(\omega)},
\end{eqnarray*}
Furthermore,
\begin{eqnarray*}
A_{kjj^{'}}&=&A_{kjj^{'}}(\omega^{'},\omega)=\int_{-1}^{+1}\,d(\cos{\vartheta})\left(\overline{S}^{kj^{'}}_{\omega^{'},-}(\vartheta)S^{kj}_{\omega,-}(\vartheta)+\overline{S}^{kj^{'}}_{\omega^{'},+}(\vartheta)S^{kj}_{\omega,+}(\vartheta)\right),\\
B_{kjj^{'}}&=&B_{kjj^{'}}(\omega^{'},\omega)=\int_{-1}^{+1}\,d(\cos{\vartheta})\sin^{2}{\vartheta}\left(\overline{S}^{kj^{'}}_{\omega^{'},-}(\vartheta)S^{kj}_{\omega,-}(\vartheta)+\overline{S}^{kj^{'}}_{\omega^{'},+}(\vartheta)S^{kj}_{\omega,+}(\vartheta)\right),
\end{eqnarray*}
\begin{eqnarray*}
\mathfrak{A}^{kjj^{'}}_{\pm}&=&\mathfrak{A}^{kjj^{'}}_{\pm}(\omega^{'},\omega)=\overline{D}_{\mp,kj^{'}}^{(1)}(\omega^{'})+D_{\mp,kj}^{(1),(\infty)}(\omega)=\pm i\left[\frac{\kappa^{'}}{2}\left(\frac{\widehat{P}_{kj^{'}}(\omega^{'})}{{\kappa^{'}}^{2}}+\frac{M^{2}m^{4}_{e}}{{\kappa^{'}}^{4}}\right)-\frac{\widehat{P}_{kj}(\omega)}{2\kappa}\right],\\
\mathfrak{B}^{kjj^{'}}_{\pm}&=&\mathfrak{B}^{kjj^{'}}_{\pm}(\omega^{'},\omega)=\overline{D}_{\mp,kj^{'}}^{(1)}(\omega^{'})+D_{\pm,kj}^{(1),(\infty)}(\omega)=\pm i\left[\frac{\kappa^{'}}{2}\left(\frac{\widehat{P}_{kj^{'}}(\omega^{'})}{{\kappa^{'}}^{2}}+\frac{M^{2}m^{4}_{e}}{{\kappa^{'}}^{4}}\right)+\frac{\widehat{P}_{kj}(\omega)}{2\kappa}\right],
\end{eqnarray*}
and $h_{\pm}^{kjj^{'}}(\omega^{'},\omega,u)$, $\widehat{h}_{\pm}^{kjj^{'}}(\omega^{'},\omega,u)$ are $\mathcal{O}(u^{-2})$ with
\[
|h_{\pm}^{kjj^{'}}(\omega^{'},\omega,u)|\leq\frac{w_{\pm}^{kjj^{'}}(\omega^{'},\omega)}{u^{2}},\quad|\widehat{h}_{\pm}^{kjj^{'}}(\omega^{'},\omega,u)|\leq\frac{\widehat{w}_{\pm}^{kjj^{'}}(\omega^{'},\omega)}{u^{2}}
\]
where
\begin{eqnarray*}
w_{\pm}^{kjj^{'}}(\omega^{'},\omega)&=&\left|\left(D_{\mp,kj}^{(2),(\infty)}(\omega)+\overline{D}_{\mp,kj^{'}}^{(2)}(\omega^{'})+\overline{D}_{\mp,kj^{'}}^{(1)}(\omega^{'})D_{\mp,kj}^{(1),(\infty)}(\omega)\right)A_{kjj^{'}}-a^{2}\overline{f}^{\infty}_{\mp}\widetilde{f}^{\infty}_{\mp}B_{kjj^{'}}\right|,\\
\widehat{w}_{\pm}^{kjj^{'}}(\omega^{'},\omega)&=&\left|\left(D_{\pm,kj}^{(2),(\infty)}(\omega)+\overline{D}_{\mp,kj^{'}}^{(2)}(\omega^{'})+\overline{D}_{\mp,kj^{'}}^{(1)}(\omega^{'})D_{\pm,kj}^{(1),(\infty)}(\omega)\right)A_{kjj^{'}}-a^{2}\overline{f}^{\infty}_{\mp}\widetilde{f}^{\infty}_{\pm}B_{kjj^{'}}\right|.
\end{eqnarray*}
Notice that $\mathfrak{A}^{kjj^{'}}_{\pm}$, $\mathfrak{B}^{kjj^{'}}_{\pm}$, $h_{\pm}^{kjj^{'}}$, $\widehat{h}_{\pm}^{kjj^{'}}\in\mathcal{C}^{\infty}(\Omega_{I}\times\Omega_{III})$. Similar formulae and considerations hold also for $\widehat{\mathfrak{A}}^{kjj^{'}}_{\pm}$, $\widehat{\mathfrak{B}}^{kjj^{'}}_{\pm}$, $\tau_{\pm}^{kjj^{'}}$, and $\widehat{\tau}_{\pm}^{kjj^{'}}$. Moreover, since the angular eigenfunctions $S^{kj^{'}}_{\omega^{'},\pm}(\vartheta)$, $S^{kj}_{\omega,\pm}(\vartheta)$ are smooth in $\omega^{'}$, and $\omega$ (Sec. II in Batic et al. (2005)), it follows that $A_{kjj^{'}}$, $B_{kjj^{'}}\in\mathcal{C}^{\infty}(\Omega_{I}\times\Omega_{III})$. Finally, by means of the H\"{o}lder inequality, and taking into account that the angular eigenfunctions are normalized it can be easily shown that $A_{kjj^{'}}$, $B_{kjj^{'}}\leq 2$. Although \eqref{FI}, and \eqref{FII} look quite complicated, we shall show in this section how to reduce \eqref{madre} to a more amenable form.\\
To determine which phase shift operator we have to introduce in order that \eqref{modificato} makes sense we can proceed as follows. Let us consider for a moment the classical definition of the wave operators asymptotically at infinity. If we proceed exactly as in the present section, we end up with an expression analogous to \eqref{madre} but without phase shift whereas \eqref{FI}, and \eqref{FII} remain unchanged. At this point we just need to analyze for $u\in[\widehat{u}_{0},+\infty)$ the long-time behavior of the integrals containing terms of zeroth order in the expansion in powers of $1/u$. Lemma~\ref{tre} suggests the following definition for the phase shift, namely
\begin{equation}\label{fasedelta}
\delta(t):=-\alpha(\omega)~\mbox{Log}\left(\frac{\kappa}{\omega}t\right),\quad \alpha(\omega):=\epsilon(\omega)\frac{Mm^{2}_{e}}{\sqrt{\omega^{2}-m^{2}_{e}}}
\end{equation}
where as in Dimock, and Kay (1986a) \mbox{Log}~t is defined for all $t\neq 0$ by $\mbox{Log}~t:=\mbox{\upshape{sgn}}(t)\log{|t|}$.
\begin{theorem} \label{THMI}
Let $W^{\pm}_{(\infty)}$ be defined as in \eqref{modificato} with phase shift $\delta(t)$ specified by \eqref{fasedelta}. Then for every $\psi_{0}^{(\infty)}\in C^{\infty}_{0}(\Omega,d\mu_{\infty})^{4}$,
\begin{eqnarray*}
\left(W^{+}_{(\infty)}\psi_{0}^{(\infty)}\right)(x)&=&2\pi\int\limits_{\sigma(H_{\infty})}\,d\omega\overline{f}^{\infty}_{-}(\omega)\widetilde{f}^{\infty}_{-}(\omega)\sum_{j\in\mathbb{Z}\backslash\{0\}}\sum_{k\in\mathbb{Z}}\psi_{\omega}^{kj}(x)\langle\psi^{kj,(\infty)}_{\omega}|\psi^{(\infty)}_{0}\rangle_{(\infty)},\\
\left(W^{-}_{(\infty)}\psi_{0}^{(\infty)}\right)(x)&=&2\pi\int\limits_{\sigma(H_{\infty})}\,d\omega\overline{f}^{\infty}_{+}(\omega)\widetilde{f}^{\infty}_{+}(\omega)\sum_{j\in\mathbb{Z}\backslash\{0\}}\sum_{k\in\mathbb{Z}}\psi_{\omega}^{kj}(x)\langle\psi^{kj,(\infty)}_{\omega}|\psi^{(\infty)}_{0}\rangle_{(\infty)}.
\end{eqnarray*}
\end{theorem}
\begin{proof}
Let us begin with the following observations. The function $g^{kj}(\omega)=\langle\psi^{kj,(\infty)}_{\omega}|\psi^{(\infty)}_{0}\rangle_{(\infty)}$ entering in \eqref{FI}, and \eqref{FII} belongs to the space of smooth rapidly decreasing functions $\mathscr{S}(\Omega_{III})$ since by going from oblate spheroidal coordinates to Cartesian coordinates $g^{kj}$ is simply the Fourier transform of smooth initial data with compact support. Moreover, for $\omega^{'}\in\Omega_{I/II}$ by means of (7), and (8) in Batic et al. (2005) it can be checked that $\mathfrak{A}^{kjj^{'}}_{\pm}$, $\mathfrak{B}^{kjj^{'}}_{\pm}$, $h_{\pm}^{kjj^{'}}$, and $\widehat{h}_{\pm}^{kjj^{'}}$ together with their first $\omega$-derivatives are polynomially bounded in $\omega$ which implies that their products with $g^{kj}$ are again Schwarzian functions. Analogous considerations hold for $\widehat{\mathfrak{A}}^{kjj^{'}}_{\pm}$, $\widehat{\mathfrak{B}}^{kjj^{'}}_{\pm}$, $\tau_{\pm}^{kjj^{'}}$, and $\widehat{\tau}_{\pm}^{kjj^{'}}$. Finally, notice that
\[
Z(\omega^{'},\omega^{'})=\frac{\omega^{'}}{\kappa^{'}},\quad X(\omega^{'},-\omega^{'})=0.
\]
Theorem~\ref{omegaII} ensures that for $\omega^{'}\in\Omega_{II}$ there is no contribution to the wave operators asymptotically at infinity. Concerning the case $\omega^{'}\in\Omega_{I}$ we apply Theorems~\ref{TEOREMA3},~\ref{TEOREMA4},~\ref{padrice}, and \ref{finalcut} to obtain
\begin{multline*}
W^{\pm}_{(\infty),II}\psi^{(\infty)}_{0}=\int\limits_{\Omega_{II}}\,d\omega^{'}\sum_{|j^{'}|\leq j^{'}_{0}}\sum_{|j|\leq j_{0}}\sum_{|k|\leq k_{0}}\psi^{kj^{'}}_{\omega^{'}}(x)\lim_{t\to \pm\infty}\sum\limits_{\pm}\int_{\widehat{u}_{0}}^{+\infty}\,du~u^{\pm i\alpha^{'}}\\
\int\limits_{\Omega_{III}}\,d\omega~Z(\omega^{'},\omega)A_{kjj^{'}}(\omega^{'},\omega)g^{kj}(\omega)\overline{f}^{\infty}_{\mp}\widetilde{f}^{\infty}_{\mp}e^{i(\omega-\omega^{'})t+\pm i(\kappa^{'}-\kappa)u+i\delta(t)}.
\end{multline*}
The result follows by applying Theorem~\ref{TEOREMA3}.\hspace{5mm}$\square$
\end{proof}

\begin{acknowledgments}
The research of D.B. was supported by the EU grant HPRN-CT-2002-00277. The author thanks Professor Felix Finster, Universit\"{a}t Regensburg, Germany, for the many hours and the thoughtful attention he dedicated to the development of this work. Special thanks go to Prof. Gian Michele Graf, Institut for Theoretical Physics, ETH Zurich, Switzerland and to Dr. Harald Schmid for their many valuable comments during numerous fruitful discussions and for their careful reading of the manuscript.
\end{acknowledgments}
\appendix

\section{\label{A:2} Oscillating integrals}
Let $\omega\in\Omega_{III}$, $\omega^{'}\in\Omega_{I}$, $\kappa$ be defined as in Section~\ref{sec:2}, $\kappa^{'}=\epsilon(\omega^{'})\sqrt{{\omega^{'}}^{2}-m^{2}_{e}}$, $\widehat{u}_{0}>0$, and $\alpha^{'}:=Mm_{e}^{2}/\kappa^{'}$. Moreover, let $\mathcal{F}\in\mathscr{S}(\Omega_{III})$ where $\mathscr{S}$ denotes the Schwarzian space of smooth rapidly decreasing functions.
\begin{lemma} \label{due}
\begin{eqnarray}
\mathcal{I}_{1,\pm}&:=&\lim_{t\to\pm\infty}\int_{\widehat{u}_{0}}^{+\infty}\,du\hspace{1mm}u^{\pm i\alpha^{'}}\int\limits_{\Omega_{III}}\,d\omega\mathcal{F}(\omega)e^{i(\omega-\omega^{'})t\pm i(\kappa^{'}-\kappa)u}=0\quad\mbox{if}\quad \mathcal{F}(+\omega^{'})=0, \label{T1} \\
\mathcal{I}_{2,\pm}&:=&\lim_{t\to\pm\infty}\int_{\widehat{u}_{0}}^{+\infty}\,du\hspace{1mm}u^{\pm i\alpha^{'}}\int\limits_{\Omega_{III}}\,d\omega \mathcal{F}(\omega)e^{i(\omega-\omega^{'})t\pm i(\kappa^{'}+\kappa)u}=0\quad\mbox{if}\quad \mathcal{F}(-\omega^{'})=0 \label{T2}
\end{eqnarray}
where the subscript $\pm$ attached to $\mathcal{I}_{1/2}$ corresponds to the $\pm$ entering in the exponents of the integrands.
\end{lemma}
\begin{proof}
Concerning $\mathcal{I}_{1,\pm}$, after the introduction of a convergence generating factor $e^{-\sigma u}$ with $\sigma>0$ we can apply Fubini theorem to obtain
\begin{equation} \label{T1sigma}
\mathcal{I}_{1,\pm}=\widehat{u}_{0}^{\pm i\alpha^{'}}\lim_{t\to\pm\infty}\lim_{\sigma\to 0^{+}}\int\limits_{\Omega_{III}}\,d\omega\hspace{1mm}\mathcal{F}(\omega)\hspace{1mm}e^{i(\omega-\omega^{'})t}\int_{1}^{+\infty}\,d\tau\hspace{1mm}\tau^{\pm i\alpha^{'}}e^{-\widetilde{z}_{\pm}\tau}
\end{equation}
with $\tau=u/\widehat{u}_{0}$, and
\begin{equation} \label{zetatilde}
\widetilde{z}_{\pm}:=\widetilde{z}_{\pm}(\omega)=[\sigma\mp i(\kappa^{'}-\kappa)]\widehat{u}_{0}.
\end{equation}
By means of 13.2.6, and 13.1.3 in Abramowitz, and Stegun we can perform the integration over $\tau$, and $\mathcal{I}_{1,\pm}$ becomes
\begin{multline} \label{T1svil}
\mathcal{I}_{1,\pm}=\widehat{u}_{0}^{1\pm i\alpha^{'}}\Gamma(1\pm i\alpha^{'})\lim_{t\to\pm\infty}\lim_{\sigma\to 0^{+}}\int\limits_{\Omega_{III}}\,d\omega\hspace{1mm}\mathcal{F}(\omega)\widetilde{z}~^{\mp i\alpha^{'}-1}\hspace{1mm}e^{i(\omega-\omega^{'})t}\\
-\frac{\widehat{u}_{0}^{1\pm i\alpha^{'}}}{1\pm i\alpha^{'}}\lim_{t\to\pm\infty}\lim_{\sigma\to 0^{+}}\int\limits_{\Omega_{III}}\,d\omega\hspace{1mm}\mathcal{F}(\omega)\hspace{1mm}e^{-\widetilde{z}_{\pm}}M(1,2\pm i\alpha^{'};\widetilde{z}_{\pm})\hspace{1mm}e^{i(\omega-\omega^{'})t}
\end{multline}
where $M$ denotes the confluent hypergeometric function of the first kind. Regarding the second integral in the above expression, let us define the function
\[
\mathcal{F}_{\sigma}(\omega):=\mathcal{F}(\omega)e^{-\widetilde{z}_{\pm}}M(1,2\pm i\alpha^{'};\widetilde{z}_{\pm})e^{i(\omega-\omega^{'})t}.
\]
Using the integral representation for the confluent hypergeometric function of the first kind 13.2.1 (ibid.), we obtain the estimate $|M(1,2\pm i\alpha^{'};\widetilde{z}_{\pm})|\leq\sqrt{1+{\alpha^{'}}^{2}}e^{\sigma\widehat{u}_{0}}$ from which it follows that $|\mathcal{F}_{\sigma}(\omega)|\leq \sqrt{1+{\alpha^{'}}^{2}}|\mathcal{F}(\omega)|$. Hence, the Lebesgue dominated convergence theorem can be applied to the second term in \eqref{T1svil} which simplifies to
\[
\lim_{t\to\pm\infty}\int\limits_{\Omega_{III}}\,d\omega\hspace{1mm}\widetilde{\mathcal{F}}(\omega)M(1,2\pm i\alpha^{'};\widehat{z}_{\pm})\hspace{1mm}e^{i(\omega-\omega^{'})t}
\]
with $\widetilde{\mathcal{F}}(\omega)=\mathcal{F}(\omega)e^{-\widehat{z}_{\pm}}$, and $\widehat{z}_{\pm}:=\mp iu_{1}(\kappa^{'}-\kappa)$. Since $\widetilde{F}(\omega)\in\mathscr{S}(\Omega_{III})$, and $M(1,2\pm i\alpha^{'};\widehat{z}_{\pm})$ is bounded by $\sqrt{1+{\alpha^{'}}^{2}}$, the Riemann-Lebesgue lemma implies that the above integral is zero for $t\to\pm\infty$, and \eqref{T1svil} becomes
\begin{equation} \label{sempsemp}
\mathcal{I}_{1,\pm}=\Gamma(1\pm i\alpha^{'})\lim_{t\to\pm\infty}\lim_{\sigma\to 0^{+}}\int\limits_{\Omega_{III}}\,d\omega\hspace{1mm}\frac{\mathcal{F}(\omega)}{[\sigma\mp i(\kappa^{'}-\kappa)]^{1\pm i\alpha^{'}}}\hspace{1mm}e^{i(\omega-\omega^{'})t}.
\end{equation}
Let us rewrite the fraction entering in the above integral as follows
\[
\frac{\mathcal{F}(\omega)}{[\sigma\mp i(\kappa^{'}-\kappa)]^{1\pm i\alpha^{'}}}=\frac{\mathcal{F}(\omega)}{\kappa-\kappa^{'}}\frac{\kappa-\kappa^{'}}{\sigma\pm i(\kappa-\kappa^{'})}[\sigma\mp i(\kappa^{'}-\kappa)]^{\mp i\alpha^{'}}.
\]
Notice that $|(\kappa-\kappa^{'})/[\sigma\pm i(\kappa-\kappa^{'})]|\leq 1$. Moreover, in the case $\epsilon(\omega^{'})=-\epsilon(\omega)$ the function $\mathcal{F}(\omega)/(\kappa-\kappa^{'})$ is integrable, and we can immediately apply the Lebesgue dominated convergence theorem to take the limit $\sigma\to 0^{+}$ inside the integral in \eqref{sempsemp} whereas for $\epsilon(\omega^{'})=\epsilon(\omega)$ we make in the above expression the substitution
\[
\frac{\mathcal{F}(\omega)}{\kappa-\kappa^{'}}=\frac{\mathcal{F}(\omega)}{\omega-\omega^{'}}\frac{\omega-\omega^{'}}{\kappa-\kappa^{'}},
\]
and observe that $\mathcal{F}(\omega)/(\omega-\omega^{'})$ has rapid decay for $|\omega|\to\infty$, and it is integrable since $\mathcal{F}(\omega^{'})=0$. Hence, \eqref{sempsemp} reduces to
\[
\mathcal{I}_{1,\pm}=\frac{\Gamma(1\pm i\alpha^{'})}{(\pm i)^{1\pm i\alpha^{'}}}\lim_{t\to\pm\infty}\int\limits_{\Omega_{III}}\,d\omega\hspace{1mm}\frac{\mathcal{F}(\omega)}{\kappa-\kappa^{'}}(\kappa-\kappa^{'})^{\mp i\alpha^{'}}e^{i(\omega-\omega^{'})t}.
\]
Finally, the Riemann-Lebesgue lemma implies that $\mathcal{I}_{1,\pm}=0$. We omit the proof for $\mathcal{I}_{2,\pm}$ since it resembles that one for $\mathcal{I}_{1,\pm}$.\hspace{5mm}$\square$
\end{proof}
\begin{lemma} \label{tre}
Let $\mathcal{I}_{1,\pm}$, and $\mathcal{I}_{2,\pm}$ be defined as in Lemma~\ref{due}. Then
\begin{eqnarray*}
\mathcal{I}_{1,+}&\sim&\left\{\begin{array}{ll}
            \frac{2\pi\kappa^{'}}{\omega^{'}}\mathcal{F}(\omega^{'})~e^{+i\alpha^{'}\log{(+t)}}&\mbox{if $t\to+\infty$}\\
            0& \mbox{if $t\to-\infty$}
            \end{array}\right.,\\
\mathcal{I}_{1,-}&\sim&\left\{\begin{array}{ll}
            0&\mbox{if $t\to+\infty$}\\
            \frac{2\pi\kappa^{'}}{\omega^{'}}\mathcal{F}(\omega^{'})~e^{-i\alpha^{'}\log{\left(-t\right)}}&\mbox{if $t\to-\infty$}
            \end{array}\right.,\\
\mathcal{I}_{2,+}&\sim&\left\{\begin{array}{ll}
            0& \hspace{5mm}\mbox{if $t\to+\infty$}\\
            \frac{2\pi\kappa^{'}}{\omega^{'}}\mathcal{F}(-\omega^{'})~e^{-2i\omega^{'}t}&\hspace{5mm}\mbox{if $t\to-\infty$}
            \end{array}\right.,\\
\mathcal{I}_{2,-}&\sim&\left\{\begin{array}{ll}
            \frac{2\pi\kappa^{'}}{\omega^{'}}\mathcal{F}(-\omega^{'})~e^{-2i\omega^{'}t}&\hspace{5mm}\mbox{if $t\to+\infty$}\\
            0& \hspace{5mm}\mbox{if $t\to-\infty$}
            \end{array}\right. .
\end{eqnarray*}
\end{lemma}
\begin{proof}
We compute $\mathcal{I}_{1,+}$. To this purpose let us rewrite $\mathcal{F}(\omega)$ as follows
\begin{equation}\label{unoz}
\mathcal{F}(\omega)=\widehat{\mathcal{F}}(\omega)+\mathcal{F}(\omega^{'})\frac{{\omega^{'}}^{2}+1}{\omega^{2}+1},\quad
\widehat{F}(\omega):=\mathcal{F}(\omega)-\mathcal{F}(\omega^{'})\frac{{\omega^{'}}^{2}+1}{\omega^{2}+1}.
\end{equation}
If we substitute \eqref{unoz} in the expression for $\mathcal{I}_{1,+}$ since $\widehat{\mathcal{F}}(\omega^{'})=0$, Lemma~\ref{due} implies that 
\[
\mathcal{I}_{1,+}=({\omega^{'}}^{2}+1)\mathcal{F}(\omega^{'})\lim_{t\to\pm\infty}\int_{\widehat{u}_{0}}^{+\infty}\,du\hspace{1mm}u^{i\alpha^{'}}\int\limits_{\Omega_{III}}\,d\omega\frac{e^{i(\omega-\omega^{'})t+i(\kappa^{'}-\kappa)u}}{\omega^{2}+1}.
\]
Notice that by introducing a convergence generating factor $e^{-\sigma u}$ with $\sigma>0$ in the above expression, we can apply the Fubini theorem, and compute the integral over $u$ exactly as in Lemma~\ref{due}. Hence, we get
\begin{multline*}
\mathcal{I}_{1,+}=\widehat{u}_{0}^{1+i\alpha^{'}}({\omega^{'}}^{2}+1)\mathcal{F}(\omega^{'})\lim_{t\to\pm\infty}\lim_{\sigma\to0^{+}}\left(\Gamma(1+i\alpha^{'})\int\limits_{\Omega_{III}}\,d\omega\frac{e^{i(\omega-\omega^{'})t}}{\widetilde{z}_{+}^{1+i\alpha^{'}}(\omega^{2}+1)}\right.\\
\left.-\frac{1}{1+i\alpha^{'}}\int\limits_{\Omega_{III}}\,d\omega\hspace{1mm}\frac{e^{-\widetilde{z}_{+}}M(1,2+i\alpha^{'};\widetilde{z}_{+})}{\omega^2+1}e^{i(\omega-\omega^{'})t}\right)
\end{multline*}
with $\widetilde{z}_{+}=[\sigma-i(\kappa^{'}-\kappa)]\widehat{u}_{0}$ but since the second integral in the above expression is a particular case of the second integral entering in \eqref{T1svil}, we can conclude that
\[
\mathcal{I}_{1,+}=\frac{\Gamma(1+i\alpha^{'})}{i^{1+i\alpha^{'}}}({\omega^{'}}^{2}+1)\mathcal{F}(\omega^{'})\lim_{t\to\pm\infty}\lim_{\sigma\to 0^{+}}\int\limits_{\Omega_{III}}\,d\omega\frac{e^{i(\omega-\omega^{'})t}}{(\omega^{2}+1)(\epsilon(\omega)\sqrt{\omega^{2}-m^{2}_{e}}-\kappa_{p})^{1+i\alpha^{'}}}
\]
with $\kappa_{p}=\kappa^{'}+i\sigma$. Since Lemma~\ref{pmm} implies that the contributions of the frequency intervals $I_{\pm}$ to the wave operators asymptotically at infinity can be made arbitrary small for $t\to\pm\infty$, we extend the domain of integration from $\Omega_{III}$ to $\sigma(H_{\infty})$ in the above integral, and compute it with the method of contour integrals. Let us define
\begin{equation} \label{esse}
\mathcal{I}_{1,+}^{(\pm)}=\lim_{t\to\pm\infty}\lim_{\sigma\to 0^{+}}\int\limits_{\sigma(H_{\infty})}\,d\omega\frac{e^{i(\omega-\omega^{'})t}}{(\omega^{2}+1)(\epsilon(\omega)\sqrt{\omega^{2}-m^{2}_{e}}-\kappa_{p})^{1+i\alpha^{'}}}.
\end{equation}
Concerning $\mathcal{I}_{1,+}^{(+)}$, we complexify the integrand as follows
\[
F(z)=\frac{e^{i(z-\omega^{'})t}}{(z^{2}+1)(\sqrt{z^{2}-m^{2}_{e}}-\kappa_{p})^{1+i\alpha^{'}}}.
\]
Since for $t\to+\infty$ the imaginary part of $e^{izt}$ decays exponentially in the complex upper half plane, we shall close the contour there. Moreover, $F(z)$ has has two simple poles at $±i$, and two branch points at $\pm m_{e}$. We make $F(z)$ single-valued by choosing $\vartheta_{1}\in[0,2\pi)$ and $\vartheta_{2}\in(-\pi,\pi]$ with $\theta_{1}:=\mbox{Arg}(z-m_{e})$, and $\vartheta_{2}:=\mbox{Arg}(z+m_{e})$ such that $\sqrt{z^2-m_{e}^{2}}=+\sqrt{\omega^2-m_{e}^{2}}$ for $\vartheta_{1}=0$, and $\sqrt{z^2-m_{e}^{2}}=-\sqrt{\omega^2-m_{e}^{2}}$ for $\vartheta_{2}=\pi$. To understand how to close the contour let us rewrite $F(z)$ as follows
\[
F(z)=\frac{e^{i(z-\omega^{'})t}}{z^{2}+1}\left(\frac{\sqrt{z^{2}-m^{2}_{e}}+\kappa_{p}}{z^{2}-m^{2}_{e}-\kappa^{2}_{p}}\right)^{1+i\alpha^{'}},
\] 
and analyze the roots of the equation $z^{2}-m^{2}_{e}-\kappa^{2}=0$. A simple calculation involving the definition of $\kappa_{p}$ gives $z^{2}={\omega^{'}}^{2}-\sigma^{2}+2i\sigma\kappa^{'}$. Notice that ${\omega^{'}}^{2}-\sigma^{2}>m^{2}_{e}-\sigma^{2}+\epsilon^{2}+2m_{e}\epsilon$ for every $\omega^{'}\in\Omega_{I}$. Since we are free to choose $\sigma$ such that $0<\sigma<m_{e}$, it follows that ${\omega^{'}}^{2}-\sigma^{2}>0$. Taking into account that the sign of $\sigma\kappa^{'}$ depends on the sign of $\omega^{'}$, we conclude that $z^{2}$ lays in the first quadrant for $\epsilon(\omega^{'})=+1$, and in the fourth quadrant for $\epsilon(\omega^{'})=-1$. Hence, it follows that $z^{2}-m^{2}_{e}-\kappa^{2}=0$ possesses for $\epsilon(\omega^{'})=+1$ two complex roots , let us say $\omega_{1,>}$, and $\omega_{2,>}=-\omega_{1,>}$, in the first, and third quadrant, respectively whereas it has for $\epsilon(\omega^{'})=-1$ a complex roots $\omega_{1,<}$ in the fourth quadrant, and another complex root $\omega_{2,<}=-\omega_{1,<}$ in the second quadrant.\\
Let us begin with the case $\epsilon(\omega^{'})=+1$. Since we want to apply the residue theorem, we choose a contour $\mathcal{C}$ such that it circumvents the point $\omega_{1,>}$, and $F(z)$ is analytic within, and on $\mathcal{C}$ except for the simple pole at $z=i$. This can be done by fixing the contour $\mathcal{C}$ as in Figure~\ref{fig0}.
\begin{figure} 
\begin{center}
\input{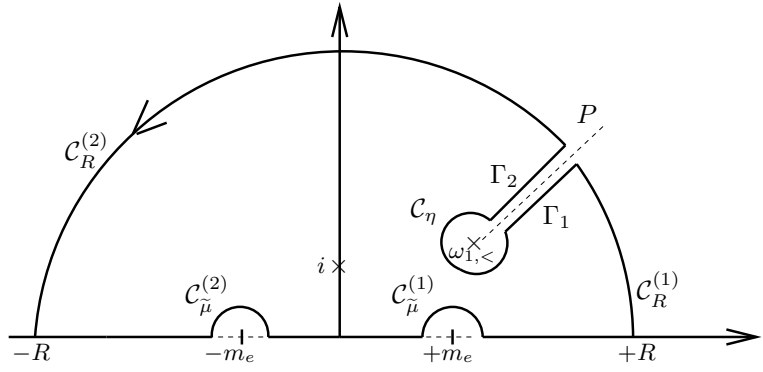}
\caption{Integration around the contour $\mathcal{C}$.}
\label{fig0}
\end{center}
\end{figure}
Let us define
\[
S_{>}=\int\limits_{\sigma(H_{\infty})}\,d\omega\frac{e^{i(\omega-|\omega^{'}|)t}}{(\omega^{2}+1)(\epsilon(\omega)\sqrt{\omega^{2}-m^{2}_{e}}-\kappa_{p,>})^{1+i\alpha^{'}_{>}}}
\]
with $\kappa_{p,>}=|\kappa^{'}|+i\sigma$, $\alpha^{'}_{>}=Mm^{2}_{e}/\sqrt{{\omega^{'}}^{2}-m^{2}_{e}}$, and $\Gamma=\Gamma_{1}\cup\mathcal{C}_{\eta}\cup\Gamma_{2}$. Then, we have
\begin{multline} \label{contorno}
S_{>}+\sum_{i=1}^{2}\left(\lim_{\widetilde{\mu}\to 0^{+}}\int\limits_{\mathcal{C}^{(i)}_{\widetilde{\mu}}}+\lim_{R\to+\infty}\int\limits_{\mathcal{C}_{R}^{(i)}}\right)\,dz\hspace{1mm}F_{>}(z)+\lim_{\widetilde{\mu}\to 0^{+}}\int_{-m_{e}+\widetilde{\mu}}^{+m_{e}-\widetilde{\mu}}\,d\omega\hspace{1mm}\mathfrak{U}_{\sigma}(\omega)+\\
\lim_{R\to+\infty}\int\limits_{\Gamma}\,dz\hspace{1mm}F_{>}(z)=2\pi i\hspace{1mm}\mathrm{Res}(F_{>}(z),z=i)
\end{multline}
with
\[
F_{>}(z)=\frac{e^{i(z-|\omega^{'}|)t}}{(z^{2}+1)(\sqrt{z^{2}-m^{2}_{e}}-\kappa_{p,>})^{1+i\alpha^{'}_{>}}},\quad \mathfrak{U}_{\sigma}(\omega)=\frac{e^{i(\omega-|\omega^{'}|)t}}{(\omega^{2}+1)(\sqrt{m^{2}_{e}-\omega^{2}}-\kappa_{p,>})^{1+i\alpha^{'}_{>}}}
\]
and $\Gamma_{3}:=[-m_{e}+\widetilde{\mu},m_{e}-\widetilde{\mu}]$. Let us first analyze the integrals on $\mathcal{C}^{(i)}_{\widetilde{\mu}}$. For $i=1$ by introducing the parameterization $z:=m+\widetilde{\mu}e^{i\vartheta_{1}}$, and taking into account that $F_{>}(z)$ is bounded on $\mathcal{C}^{(1)}_{\widetilde{\mu}}$ it can be easily verified that
\[
\Bigl\lvert\int\limits_{\mathcal{C}^{(1)}_{\widetilde{\mu}}}\,dz\hspace{1mm}F_{>}(z)\Bigr\rvert\leq c\widetilde{\mu}\int_{0}^{\pi}\,d\vartheta_{1}\hspace{1mm}e^{-\widetilde{\mu}t\sin{\vartheta_{1}}}\leq\pi c\widetilde{\mu}.
\]
An analogous relation holds for the integral on $\mathcal{C}^{(2)}_{\widetilde{\mu}}$. We consider now the integral on $\mathcal{C}^{(1)}_{R}$. To this purpose let us introduce the parameterization $z=Re^{i\vartheta}$ with $0\leq\vartheta\leq\vartheta_{0}-\widetilde{\epsilon}$, and $0<\widetilde{\epsilon}<\vartheta_{0}<\pi/2$ where $\vartheta_{0}$ denotes the slope of the ray $P$ in Figure~\ref{fig0}. Since $F_{>}(z)$ is bounded on $\mathcal{C}^{(1)}_{R}$, we have
\[
\Bigl\lvert\int\limits_{\mathcal{C}^{(1)}_{R}}\,dz\hspace{1mm}F(z)\Bigr\rvert\leq C\int_{0}^{\vartheta_{0}-\widetilde{\epsilon}}\,d\vartheta\hspace{1mm}e^{-Rt\sin{\vartheta}}  
\]
For $R\to+\infty$ the above integral converges to zero according to Lebesgue dominated convergence theorem, and the integral on $\mathcal{C}^{(2)}_{R}$ can be treated analogously. Regarding the third term in \eqref{contorno} we find for $0<\sigma<m_{e}$
\[
\Bigl\lvert\int_{-m_{e}+\widetilde{\mu}}^{+m_{e}-\widetilde{\mu}}\,d\omega\hspace{1mm}\mathfrak{U}_{\sigma}(\omega)\Bigr\rvert\leq\int_{-m_{e}+\widetilde{\mu}}^{+m_{e}-\widetilde{\mu}}\,\frac{d\omega}{\sqrt{(\sqrt{m^{2}_{e}-\omega^{2}}-|\kappa^{'}|)^{2}+\sigma^{2}}}\leq\frac{2}{\sigma}(m_{e}-\widetilde{\mu}),
\]
and by taking the limit $\widetilde{\mu}\to 0^{+}$ we obtain
\[
\lim_{\widetilde{\mu}\to 0^{+}}\int_{-m_{e}+\widetilde{\mu}}^{+m_{e}-\widetilde{\mu}}\,d\omega\hspace{1mm}\mathfrak{U}_{\sigma}(\omega)=\int_{-m_{e}}^{+m_{e}}\,d\omega\hspace{1mm}\mathfrak{U}_{\sigma}(\omega).
\]
We consider now
\begin{equation}\label{telesette}
\lim_{t\to\pm\infty}\lim_{\sigma\to 0^{+}}\int_{-m_{e}}^{+m_{e}}\,d\omega\hspace{1mm}\mathfrak{U}_{\sigma}(\omega).
\end{equation}
Since $|\mathfrak{U}_{\sigma}(\omega)|\leq 1/\sqrt{(\sqrt{m^{2}_{e}-\omega^{2}}-|\kappa^{'}|)^{2}+\sigma^{2}}$, and
\[
\sqrt{m^{2}_{e}-\omega^{2}}-|\kappa^{'}|=\sqrt{m^{2}_{e}-\omega^{2}}-\sqrt{{\omega^{'}}^{2}-m^{2}_{e}}\geq \sqrt{m^{2}_{e}-\omega^{2}}-\sqrt{\omega_{0}^{2}-m^{2}_{e}}\geq-\sqrt{\omega_{0}^{2}-m^{2}_{e}},
\]
it follows that $|\mathfrak{U}_{\sigma}(\omega)|\leq 1/\sqrt{\omega_{0}^{2}-m^{2}_{e}}$. Hence, by applying the Lebesgue dominated convergence theorem to \eqref{telesette} we end up with 
\begin{equation}\label{teleotto}
\lim_{t\to\pm\infty}\int_{-m_{e}}^{+m_{e}}\,d\omega\hspace{1mm}\frac{f(\omega)e^{i(\omega-|\omega^{'}|)t}}{(\omega^{2}+1)(\sqrt{m_{e}^{2}-\omega^{2}}-|\kappa^{'}|)^{i\alpha^{'}_{>}}},\quad f(\omega)=\frac{\sqrt{m_{e}^{2}-\omega^{2}}+|\kappa^{'}|}{2m^{2}_{e}-{\omega^{'}}^{2}-\omega^{2}}.
\end{equation}
Since $2m^{2}_{e}-{\omega^{'}}^{2}-\omega^{2}\geq m^{2}_{e}-{\omega^{'}}^{2}$ for $|\omega|\leq m_{e}$ it results that
\[
f(\omega)\leq\frac{\sqrt{m_{e}^{2}-\omega^{2}}+|\kappa^{'}|}{m_{e}^{2}-{\omega^{'}}^{2}},\quad\omega^{'}\in\Omega_{I}
\]
is integrable, and the Riemann-Lebesgue lemma implies that \eqref{teleotto} is zero. Regarding the computation of the residue at $z=+i$, it can be checked that
\[
|\mbox{Res}(F_{>}(z),z=+i)|=\frac{e^{-t}}{2||\kappa^{'}|+i(\sigma-\sqrt{1+m^{2}_{e}})|}\leq\frac{e^{-t}}{2|\kappa^{'}|},\quad |\kappa^{'}|\quad\mbox{for}\quad \omega^{'}\in\Omega_{I}.
\]
Therefore, it follows that it does not give any contribution for $t\to+\infty$. Finally, since for $R\to\infty$ the contour $-\Gamma$ goes over into the Gamma function contour $\Gamma^{'}$ \eqref{contorno} simplifies to
\begin{equation} \label{plenaria}
\lim_{t\to+\infty}\lim_{\sigma\to0^{+}}S_{>}=\lim_{t\to+\infty}\lim_{\sigma\to0^{+}}\int\limits_{\Gamma^{'}}\,dz\hspace{1mm}F_{>}(z).
\end{equation}
We reduce now the complex integral in \eqref{plenaria} to the Hankel contour integral for the reciprocal Gamma function. To this purpose let us rewrite $F_{>}(z)$ as follows
\[
F_{>}(z)=e^{i(\omega_{1,>}-|\omega^{'}|)t}\widetilde{F}_{>}(z)e^{i(z-\omega_{1,>})t}(z-\omega_{1,>})^{\beta-1}
\]
with
\[
\widetilde{F}_{>}(z)=\frac{1}{z^2+1}\left(\frac{\sqrt{z^2-m^2}+\kappa_{p,>}}{z+\omega_{1,>}}\right)^{1-\beta},\quad\beta=-i\alpha^{'}_{>}. 
\]
Since $\widetilde{F}_{>}(z)$ is continuous on, and analytic within $\Gamma^{'}$, the integral in \eqref{plenaria} is convergent, and $\widetilde{F}_{>}(z)$ admits a convergent expansion around $\omega_{1,>}$, namely
\[
\widetilde{F}_{>}(z)=\sum_{n=0}^{\infty}c_{n}(z-\omega_{1,>})^{n}\quad |z-\omega_{1,>}|<\eta,\quad\eta>0.
\] 
Thus,
\[
\int\limits_{\Gamma^{'}}\,dz\hspace{1mm}F_{>}(z)=e^{i(\omega_{1,>}-|\omega^{'}|)t}\sum_{n=0}^{\infty}c_{n}\int\limits_{\Gamma^{'}}\,dz\hspace{1mm}(z-\omega_{1,>})^{\beta+n-1}\hspace{1mm}e^{i(z-\omega_{1,>})t}.
\]
A simple calculation employing the Hankel contour integral of the reciprocal Gamma function (see Erd$\acute{e}$ly et al., Vol.I, p.14) gives
\[
\int\limits_{\Gamma^{'}}\,dz\hspace{1mm}F_{>}(z)=2\pi i^{1+i\alpha_{>}^{'}} e^{i(\omega_{1,>}-|\omega^{'}|)t}e^{i\alpha_{>}^{'}\log{t}}\sum_{n=0}^{\infty}\frac{c_{n}}{i^{n}\Gamma(1+i\alpha_{>}^{'}-n)}t^{-n}.
\]
Since the coefficients $c_{n}$ depend analytically on $\sigma$, we can perform the limit $\sigma\to 0^{+}$, and we obtain
\[
\lim_{\sigma\to 0^{+}}S_{>}\sim\frac{2\pi i^{1+i\alpha_{>}^{'}}}{\Gamma(1+i\alpha_{>}^{'})}\frac{1}{{\omega^{'}}^{2}+1}\left(\frac{\kappa^{'}}{\omega^{'}}\right)^{1+i\alpha^{'}_{>}}e^{i\alpha_{>}^{'}\log{t}},\quad t\to+\infty.
\]
Taking into account that $\omega^{'}\in\Omega_{I}$, we have
\begin{equation}\label{telenove}
\lim_{\sigma\to 0^{+}}S_{>}\sim\frac{2\pi i^{1+i\alpha_{>}^{'}}}{\Gamma(1+i\alpha_{>}^{'})}\frac{\kappa^{'}}{{\omega^{'}(\omega^{'}}^{2}+1)}~e^{i\alpha_{>}^{'}\log{t}},\quad t\to+\infty.
\end{equation}
We consider now the case $\epsilon(\omega^{'})=-1$. In analogy to $S_{>}$ we define
\[
S_{<}:=\int\limits_{\sigma(H_{\infty})}\,d\omega\frac{e^{i(\omega+|\omega^{'}|)t}}{(\omega^{2}+1)(\epsilon(\omega)\sqrt{\omega^{2}-m^{2}_{e}}-\kappa_{p,<})^{1+i\alpha^{'}_{<}}}
\]
with $\kappa_{p,<}=-|\kappa^{'}|+i\sigma$, $\alpha^{'}_{<}=-Mm^{2}_{e}/\sqrt{{\omega^{'}}^{2}-m^{2}_{e}}$, and
\[
F_{<}(z)=\frac{e^{i(z+|\omega^{'}|)t}}{(z^{2}+1)(\sqrt{z^{2}-m^{2}_{e}}-\kappa_{p,<})^{1+i\alpha^{'}_{<}}}.
\]
Proceeding as we did for $S_{>}$, and taking into account that $\omega_{1,<}$ lays on the second quadrant, we end up with
\begin{equation}\label{teledieci}
\lim_{\sigma\to 0^{+}}S_{<}\sim\frac{2\pi i^{1+i\alpha_{<}^{'}}}{\Gamma(1+i\alpha_{<}^{'})}\frac{1}{{\omega^{'}}^{2}+1}\left(\frac{\kappa^{'}}{\omega^{'}}\right)e^{i\alpha_{<}^{'}\log{t}},,\quad t\to+\infty.
\end{equation}
Putting together \eqref{telenove}, and \eqref{teledieci} and taking into account that $\omega^{'}\in\Omega_{I}$, we obtain for $t\to+\infty$
\[
\lim_{\sigma\to 0^{+}}\int\limits_{\sigma(H_{\infty})}\,d\omega\frac{e^{i(\omega-\omega^{'})t}}{(\omega^{2}+1)(\epsilon(\omega)\sqrt{\omega^{2}-m^{2}_{e}}-\kappa_{p})^{1+i\alpha^{'}}}\sim\frac{2\pi i^{1+i\alpha^{'}}}{\Gamma(1+i\alpha^{'})}\frac{\kappa^{'}}{\omega^{'}({\omega^{'}}^{2}+1)}~e^{i\alpha^{'}\log{t}},
\]
from which it results that
\[
\mathcal{I}_{1,+}\sim \frac{2\pi\kappa^{'}}{\omega^{'}}\mathcal{F}(\omega^{'})~e^{i\alpha^{'}\log{(+t)}}. 
\]
Let us consider the case $t\to-\infty$. We complexify the integrand function in \eqref{esse} as follows
\[
\widehat{F}(z)=\frac{e^{i(z-\omega^{'})t}}{(z^2+1)(-\sqrt{z^2-m_{e}^{2}}-\kappa_{p})^{1+i\alpha^{'}}}.
\]
Since for $t\to-\infty$ the imaginary part of $e^{izt}$ decays exponentially for $\mbox{Im}~z<0$, we have to choose the contour in the complex lower half-plane. Moreover, we recall that $\sqrt{z^2-m_{e}^{2}}=-\sqrt{\omega^{2}-m_{e}^{2}}$ for $\vartheta_{1}=2\pi$, and $\sqrt{z^2-m_{e}^{2}}=+\sqrt{\omega^{2}-m_{e}^{2}}$ for $\vartheta_{2}=-\pi$. In what follows we outline the computation in the case $\epsilon(\omega^{'})=+1$ since the case $\epsilon(\omega^{'})=-1$ is similar. In order to apply the residue theorem we fix the contour $\widetilde{\mathcal{C}}$ such that it circumvents the point $-\omega_{1,>}$. Moreover, the function 
\[
\widehat{F}_{>}(z)=\frac{e^{i(z-|\omega^{'}|)t}}{(z^2+1)(-\sqrt{z^2-m_{e}^{2}}-\kappa_{p,>})^{1+i\alpha^{'}_{>}}}
\] 
is analytic within, and on $\widetilde{\mathcal{C}}$ except for the simple pole at $-i$. The residue theorem implies that
\[
\int\limits_{\widetilde{\mathcal{C}}}\,dz\hspace{1mm}\widehat{F}_{>}(z)=-2\pi i\mbox{Res}(\widehat{F}_{>}(z),z=-i).
\]
A simple computation shows that the residue is dominated by $e^{t}$. At this point we can proceed similarly as we did for \eqref{contorno} with the only difference that now the part of $\widehat{F}_{>}(z)$ which is continuous on, and analytic within that part of the contour indentating the point $-\omega_{1,>}$, admits for $|z+\omega_{1,>}|<\eta$ with $\eta>0$ a convergent expansion $\sum_{n=0}^{\infty}\widehat{c}_{n}(z+\omega_{1,>})^{n}$ such that $\widehat{c}_{0}$ tends to zero for $\sigma\to 0^{+}$. Repeating the same procedure for $\epsilon(\omega^{'})=-1$, we conclude that $\mathcal{I}_{1,+}\sim 0$ for $t\to-\infty$. Finally, $\mathcal{I}_{1,-}$ can be obtained directly from $\mathcal{I}_{1,+}$ by means of complex conjugation, and of the transformation $t\to-t$. $\mathcal{I}_{2,-}$ can be computed by the same method used for $\mathcal{I}_{1,+}$ whereas $\mathcal{I}_{2,+}$ can be derived from $\mathcal{I}_{2,-}$ by complex conjugation, and the transformation $t\to-t$.\hspace{5mm}$\square$
\end{proof}

\section{\label{A:3} Theorems for the evaluation of $W^{\pm}_{(\infty)}$}
Let $\omega\in\Omega_{III}$, $\omega^{'}\in\Omega_{I}$, $\kappa$ be defined as in Section~\ref{sec:2}, $\kappa^{'}=\epsilon(\omega^{'})\sqrt{{\omega^{'}}^{2}-m^{2}_{e}}$, $\widehat{u}_{0}>0$, and $\alpha^{'}:=Mm_{e}^{2}/\kappa^{'}$. Moreover, let $\mathcal{F}\in\mathscr{S}(\Omega_{III})$, $\delta(t)$ be given by \eqref{fasedelta}, and
\[
\widetilde{\delta}(t):=-\alpha^{'}\mbox{Log}~t,\quad \alpha^{'}:=\alpha(\omega^{'})=\epsilon(\omega^{'})\frac{Mm^{2}_{e}}{\sqrt{{\omega^{'}}^{2}-m^{2}_{e}}}.
\]
\begin{lemma} \label{stima_delta}
Let $I=I_{<}:=[\omega,\omega^{'}]$ for $\omega<\omega^{'}$, and $I=I_{>}:=[\omega^{'},\omega]$ for $\omega^{'}<\omega$, Then, 
\[
\left|\frac{e^{i\delta(t)}-e^{i\widetilde{\delta}(t)}}{\omega-\omega^{'}}\right|\leq Mm^{2}_{e}\log|t|\sup_{\omega\in I}{\{\rho(\omega)\}},\quad\rho(\omega)=\frac{|\omega|}{(\omega^{2}-m_{e}^{2})^{3/2}}.
\]
\end{lemma}
\begin{proof}
The result follows directly from the inequality
\[
\left|e^{i\delta(t)}-e^{i\widetilde{\delta}(t)}\right|=\left|\int_{\omega}^{\omega^{'}}\,dx\frac{de^{i\delta(t)}}{dx}\right| \leq|\omega-\omega^{'}|\sup_{\omega\in I}{\left\{\left|\frac{de^{i\delta(t)}}{d\omega}\right|\right\}},
\]
together with the estimate
\[
\left|\frac{de^{i\delta(t)}}{d\omega}\right|\leq Mm_{e}^{2}\log{|t|}\rho(\omega),\quad \rho(\omega)=\frac{|\omega|}{(\omega^{2}-m_{e}^{2})^{3/2}}.
\]\hspace{5mm}$\square$
\end{proof}

\begin{lemma} \label{TEOREMA2}

\begin{eqnarray*}
\Delta_{\pm}&=&\lim_{t\to\pm\infty}\int_{\widehat{u}_{0}}^{+\infty}\,du\hspace{1mm}u^{\pm i\alpha^{'}}\int\limits_{\Omega_{III}}\,d\omega\mathcal{F}(\omega)\left(e^{i\delta(t)}-e^{i\widetilde{\delta}(t)}\right)e^{i(\omega-\omega^{'})t\pm i(\kappa^{'}-\kappa)u}=0,\\
P_{\pm}&=&\lim_{t\to\pm\infty}\int_{\widehat{u}_{0}}^{+\infty}\,du\hspace{1mm}u^{\pm i\alpha^{'}}\int\limits_{\Omega_{III}}\,d\omega\mathcal{F}(\omega)\left(e^{i\delta(t)}-e^{i\widetilde{\delta}(t)}\right)e^{i(\omega-\omega^{'})t\pm i(\kappa^{'}+\kappa)u}=0
\end{eqnarray*}
where the subscript $\pm$ attached to $\Delta$, and $P$ corresponds to the $\pm$ entering in the exponents of the integrands.
\end{lemma}
\begin{proof}
We show the result for $\Delta_{\pm}$ since $P_{\pm}$ can be computed with the same method. Proceeding as in Lemma~\ref{due} we obtain
\begin{multline} \label{TT1svil}
\Delta_{\pm}=\widehat{u}_{0}^{1\pm i\alpha^{'}}\Gamma(1\pm i\alpha^{'})\lim_{t\to\pm\infty}\lim_{\sigma\to0^{+}}\int\limits_{\Omega_{III}}\,d\omega\hspace{1mm}\mathcal{F}_{(t)}(\omega)\widetilde{z}_{\pm}^{\hspace{1mm}\mp i\alpha^{'}-1}e^{i(\omega-\omega^{'})t}\\
-\frac{\widehat{u}_{0}^{1\pm i\alpha^{'}}}{1\pm i\alpha^{'}}\lim_{t\to\pm\infty}\lim_{\sigma\to0^{+}}\int\limits_{\Omega_{III}}\,d\omega\hspace{1mm}\mathcal{F}_{(t)}(\omega)e^{-\widetilde{z}_{\pm}}M(1,2\pm i\alpha^{'};\widetilde{z}_{\pm})e^{i(\omega-\omega^{'})t}
\end{multline}
with $\mathcal{F}_{(t)}(\omega)=\mathcal{F}(\omega)(e^{i\delta(t)}-e^{i\widetilde{\delta}(t)})$. Let us consider the first integral in \eqref{TT1svil}. We rewrite the integrand as follows
\[
\mathcal{F}_{(t)}(\omega)\widetilde{z}_{\pm}^{\hspace{1mm}\mp i\alpha^{'}-1}e^{i(\omega-\omega^{'})t}=\frac{\mathcal{F}_{(t)}(\omega)}{\omega-\omega^{'}}\frac{\omega-\omega^{'}}{\kappa-\kappa^{'}}\frac{\kappa-\kappa^{'}}{\sigma\mp i(\kappa^{'}-\kappa)}\frac{e^{i(\omega-\omega^{'})t}}{[\sigma\mp i(\kappa^{'}-\kappa)]^{\pm i\alpha^{'}}}
\]
where we have used the definition of $\widetilde{z}_{\pm}$ given in Lemma~\ref{due}. Since $\mathcal{F}_{(t)}(\omega)/(\omega-\omega^{'})$ is continuous at $\omega=\omega^{'}$, and has rapid decay for $|\omega|\to\infty$, we can apply the Lebesgue dominated convergence theorem to obtain
\begin{equation} \label{referenza1}
\Delta^{(1)}:=\lim_{t\to\pm\infty}\int\limits_{\Omega_{III}}\,d\omega\hspace{1mm}\mathcal{F}_{(t)}(\omega)(\kappa-\kappa^{'})^{\mp i\alpha^{'}-1}e^{i(\omega-\omega^{'})t}.
\end{equation}
Without loss of generality let us suppose that $\epsilon(\omega^{'})=+1$. For $\widetilde{\epsilon}(t)=1/\log^{2}|t|$ with $|t|> e^{1/\sqrt{\omega_{0}-m_{e}-\widetilde{\mu}}}$ which ensures that $|\omega^{'}|-\widetilde{\epsilon}>m_{e}+\widetilde{\mu}$ we introduce the following decomposition of $\Omega_{III}$, namely
\[
\Omega_{III}=(-\infty,-m_{e}-\widetilde{\mu}]\cup[m_{e}+\widetilde{\mu},|\omega^{'}|-\widetilde{\epsilon}(t)]\cup[|\omega^{'}|-\widetilde{\epsilon}(t),|\omega^{'}|+\widetilde{\epsilon}(t)]\cup[|\omega^{'}|+\widetilde{\epsilon}(t),+\infty).
\]
Let us rewrite  $e^{i(\omega-\omega^{'})t}$ as follows 
\begin{equation} \label{IDENTITA1}
e^{i(\omega-\omega^{'})t}=\frac{1}{it}\frac{d}{d\omega}e^{i(\omega-\omega^{'})t}.
\end{equation}
Concerning the interval $I:=(-\infty,-m_{e}-\widetilde{\mu}]$ we use \eqref{IDENTITA1} to integrate \eqref{referenza1} by parts. In the limit $t\to\pm\infty$ we have no boundary terms since $\mathcal{F}$ has rapid decay for $|\omega|\to\infty$, and at $\omega=-m_{e}-\widetilde{\mu}$ the corresponding boundary term is dominated by $t^{-1}$. Hence we get
\[
\Delta^{(1)}_{I}=(-)^{1\mp i\alpha^{'}}\lim_{t\to\pm\infty}\frac{1}{it}\int\limits_{I}\,d\omega\frac{d}{d\omega}\left(\mathcal{F}_{(t)}(\omega)\left(\sqrt{\omega^{2}-m_{e}^{2}}+|\kappa^{'}|\right)^{\mp i\alpha^{'}-1}\right)e^{i(\omega-|\omega^{'}|)t}.
\]
Computing the derivative in the above integral, we find the following estimate
\[
|\Delta^{(1)}_{I}|\leq\lim_{t\to\pm\infty}\frac{C_{1}+C_{2}\log|t|}{|t|}
\]
with constants $C_{1}$, $C_{2}>0$. Thus, $\Delta^{(1)}_{I}=0$ in the limit $t\to\pm\infty$. Regarding the interval $II:=[m_{e}+\widetilde{\mu},|\omega^{'}|-\widetilde{\epsilon}(t)]$, we use again \eqref{IDENTITA1}. The boundary terms vanish for $t\to\pm\infty$ since the corresponding boundary term at $\omega=m_{e}+\widetilde{\mu}$is dominated by $t^{-1}$ whereas at $\omega=|\omega^{'}|-\widetilde{\epsilon}(t)$ if we define the function
\[
\mathcal{G}_{(t)}(\omega):=\mathcal{F}_{(t)}(\omega)(\sqrt{\omega^{2}-m_{e}^{2}}-|\kappa^{'}|)^{\mp i\alpha^{'}-1}e^{i(\omega-|\omega^{'}|)t},
\]
we find that
\[
\lim_{t\to\pm\infty}\left|\frac{\mathcal{G}_{(t)}(|\omega^{'}|-\widetilde{\epsilon}(t))}{t}\right|\leq 2\frac{\kappa^{'}}{\omega^{'}}|\mathcal{F}(\omega^{'})|\lim_{t\to\pm\infty}\frac{\log^{2}{|t|}}{|t|}.
\]
Thus, we obtain
\[
\Delta^{(1)}_{II}=\lim_{t\to\pm\infty}\frac{1}{it}\int\limits_{II}\,d\omega\frac{d}{d\omega}\left(\mathcal{F}_{(t)}(\omega)\left(\sqrt{\omega^{2}-m_{e}^{2}}-|\kappa^{'}|\right)^{\mp i\alpha^{'}-1}\right)e^{i(\omega-|\omega^{'}|)t}.
\]
By computing the derivative in the above expression, and by applying Lemma~\ref{stima_delta} it can be checked that
\[
|\Delta^{(1)}_{II}|\leq C_{3}\lim_{t\to\pm\infty}\frac{\log^{4}|t|}{|t|}
\]
for some constant $C_{3}>0$. Concerning the interval $III:=[|\omega^{'}|-\widetilde{\epsilon}(t),|\omega^{'}|+\widetilde{\epsilon}(t)]$ let 
\[
\Delta^{(1)}_{III}=\lim_{t\to\pm\infty}\int\limits_{III}\,d\omega\hspace{1mm}\mathcal{F}_{(t)}(\omega)(\sqrt{\omega^{2}-m_{e}^{2}}-|\kappa^{'}|)^{\mp i\alpha^{'}-1}e^{i(\omega-|\omega^{'}|)t}.
\]
By rewriting the integrand as follows
\[
\mathcal{F}_{(t)}(\omega)(\sqrt{\omega^{2}-m_{e}^{2}}-|\kappa^{'}|)^{\mp i\alpha^{'}-1}=\mathcal{F}(\omega)\frac{e^{i\delta(t)}-e^{i\widetilde{\delta}(t)}}{\omega-|\omega^{'}|}\frac{\omega-|\omega^{'}|}{\sqrt{\omega^{2}-m_{e}^{2}}-|\kappa^{'}|}(\sqrt{\omega^{2}-m_{e}^{2}}-|\kappa^{'}|)^{\mp i\alpha^{'}},
\]
and making use of Lemma~\ref{stima_delta}, we find that
\[
|\Delta^{(1)}_{III}|\leq C_{4}\lim_{t\to\pm\infty}\frac{1}{\log{|t|}}
\]
for some constant $C_{4}>0$. Finally, for $IV:=[|\omega^{'}|+\widetilde{\epsilon}(t),+\infty)$ we define
\[
\Delta^{(1)}_{IV}=\lim_{t\to\pm\infty}\int\limits_{IV}\,d\omega\hspace{1mm}\mathcal{F}_{(t)}(\omega)(\sqrt{\omega^{2}-m_{e}^{2}}-|\kappa^{'}|)^{\mp i\alpha^{'}-1}e^{i(\omega-|\omega^{'}|)t}.
\]
Employing the methods used to compute $\Delta^{(1)}_{I}$, and $\Delta^{(1)}_{II}$ it can be shown that $\Delta^{(1)}_{IV}=0$. Hence, we can conclude that $\Delta^{(1)}=0$. Concerning the second integral entering in \eqref{TT1svil} since we already showed in Lemma~\ref{due} the boundedness of the function $M(1,2\pm i\alpha^{'};\widetilde{z}_{\pm})$, we can apply the Lebesgue dominated convergence theorem to take the limit $\sigma\to 0^{+}$ inside the integral, and we end up with the following expression
\begin{equation}\label{SSEMPLICE}
\Delta^{(2)}:=\lim_{t\to\pm\infty}\int\limits_{\Omega_{III}}\,d\omega\hspace{1mm}\mathcal{F}_{(t)}(\omega)M(1,2\pm i\alpha^{'};z_{\pm})~e^{i(\omega-\omega^{'})t-z_{\pm}}
\end{equation}
with $z_{\pm}=\mp i\widehat{u}_{0}(\kappa^{'}-\kappa)$. If we now use \eqref{IDENTITA1} to integrate \eqref{SSEMPLICE} by parts we obtain
\begin{equation} \label{neviga3}
\Delta^{(2)}=-\lim_{t\to\pm\infty}\frac{1}{it}\int\limits_{\Omega_{III}}\,d\omega\frac{d}{d\omega}\left(\mathcal{F}_{(t)}(\omega)M(1,2\pm i\alpha^{'};z_{\pm})~e^{-z_{\pm}}\right)e^{i(\omega-\omega^{'})t}.
\end{equation}
We do not get any boundary term since $\mathcal{F}(\omega)$ decays rapidly for $|\omega|\to\infty$ whereas at $\omega=\pm(m_{e}+\widetilde{\mu})$ the corresponding boundary terms are dominated by $t^{-1}$. Taking into account that 13.4.12 in Abramowitz, and Stegun implies that
\[
\frac{d}{d\omega}M(1,2\pm i\alpha^{'};z_{\pm})=\pm(i\widehat{u}_{0})^{-1}\left(\frac{\kappa}{\omega}\right)\left[M(1,2\pm i\alpha^{'};z_{\pm})-\frac{1\pm i\alpha^{'}}{2\pm i\alpha^{'}}M(1,3\pm i\alpha^{'};z_{\pm})\right]
\]
with $M(1,3\pm i\alpha^{'};z_{\pm})\leq\sqrt{4+{\alpha^{'}}^{2}}$ where the bound has been obtained by means of 13.2.1 (ibid.), we can compute the derivative in \eqref{neviga3}, and we get the following estimate
\[
|\Delta^{(2)}|\leq\lim_{t\to\pm\infty}\frac{C_{5}+C_{6}\log|t|}{|t|}
\]
for some constants $C_{5}$, $C_{6}>0$. Hence, $\Delta^{(2)}=0$, and the proof is completed.\hspace{5mm}$\square$
\end{proof}

\begin{theorem} \label{TEOREMA3}
Let
\begin{eqnarray*}
I_{1}&=&\lim_{t\to\pm\infty}\int_{\widehat{u}_{0}}^{+\infty}\,du\hspace{1mm}u^{+i\alpha^{'}}\int\limits_{\Omega_{III}}\,d\omega\mathcal{F}(\omega)~e^{i\varphi^{+}_{(t)}(\omega,u)+i\delta(t)},\\
I_{2}&=&\lim_{t\to\pm\infty}\int_{\widehat{u}_{0}}^{+\infty}\,du\hspace{1mm}u^{-i\alpha^{'}}\int\limits_{\Omega_{III}}\,d\omega\mathcal{F}(\omega)~e^{i\varphi^{-}_{(t)}(\omega,u)+i\delta(t)},\\
I_{3}&=&\lim_{t\to\pm\infty}\int_{\widehat{u}_{0}}^{+\infty}\,du\hspace{1mm}u^{+i\alpha^{'}}\int\limits_{\Omega_{III}}\,d\omega\mathcal{F}(\omega)~e^{i\widehat{\varphi}^{+}_{(t)}(\omega,u)+i\delta(t)},\\
I_{4}&=&\lim_{t\to\pm\infty}\int_{\widehat{u}_{0}}^{+\infty}\,du\hspace{1mm}u^{-i\alpha^{'}}\int\limits_{\Omega_{III}}\,d\omega\mathcal{F}(\omega)~e^{i\widehat{\varphi}^{-}_{(t)}(\omega,u)+i\delta(t)}
\end{eqnarray*}
with
\[
\varphi^{\pm}_{(t)}(\omega,u)=(\omega-\omega^{'})t\pm(\kappa^{'}-\kappa)u,\quad \widehat{\varphi}^{\pm}_{(t)}(\omega,u)=(\omega-\omega^{'})t\pm(\kappa^{'}+\kappa)u.
\]
Then
\begin{eqnarray*}
I_{1}&\sim&\left\{\begin{array}{ll}
            2\pi\frac{\kappa^{'}}{\omega^{'}}\mathcal{F}(\omega^{'})
&\hspace{2cm}\mbox{if $t\to+\infty$}\\
            0&\hspace{2cm} \mbox{if $t\to-\infty$}
            \end{array}\right.,\\
I_{2}&\sim&\left\{\begin{array}{ll}
            0&\hspace{2cm} \mbox{if $t\to+\infty$}\\
            2\pi\frac{\kappa^{'}}{\omega^{'}}\mathcal{F}(\omega^{'})
& \hspace{2cm}\mbox{if $t\to-\infty$}
            \end{array}\right.,\\
I_{3}&\sim&\left\{\begin{array}{ll}
            0& \hspace{5mm}\mbox{if $t\to+\infty$}\\
            2\pi\frac{\kappa^{'}}{\omega^{'}}\mathcal{F}(-\omega^{'})~e^{-2i\omega^{'}t}& \hspace{5mm}\mbox{if $t\to-\infty$}
            \end{array}\right.,\\
I_{4}&\sim&\left\{\begin{array}{ll}
            2\pi\frac{\kappa^{'}}{\omega^{'}}\mathcal{F}(-\omega^{'})~e^{-2i\omega^{'}t}&\hspace{5mm}\mbox{if $t\to+\infty$}\\
            0& \hspace{5mm}\mbox{if $t\to-\infty$}
            \end{array}\right. .
\end{eqnarray*}
\end{theorem}
\begin{proof}
We give the proof for the first result, the others being similar. By adding, and subtracting to $e^{i\delta(t)}$ the term $e^{i\widetilde{\delta}(t)}$ we obtain
\begin{multline*}
I_{1}=\lim_{t\to\pm\infty}\int_{\widehat{u}_{0}}^{+\infty}\,du\hspace{1mm}u^{i\alpha^{'}}\int\limits_{\Omega_{III}}\,d\omega\hspace{1mm}\mathcal{F}(\omega)\left(e^{i\delta(t)}-e^{i\widetilde{\delta}(t)}\right)~e^{i\varphi^{+}_{(t)}(\omega,u)}\\
+\lim_{t\to\pm\infty}e^{i\widetilde{\delta}(t)}\int_{\widehat{u}_{0}}^{+\infty}\,du\hspace{1mm}u^{i\alpha^{'}}\int\limits_{\Omega_{III}}\,d\omega\hspace{1mm}\mathcal{F}(\omega)~e^{i\varphi^{+}_{(t)}(\omega,u)}.
\end{multline*}
The result follows by applying Lemma~\ref{TEOREMA2}, and \ref{tre} to the first term on the r.h.s. of the above expression and to the second term, respectively.\hspace{5mm}$\square$
\end{proof}

\begin{theorem} \label{TEOREMA4}
Let $\varphi^{\pm}_{(t)}(\omega,u)$, and $\widehat{\varphi}^{\pm}_{(t)}(\omega,u)$ be defined as in Theorem~\ref{TEOREMA3}, Then 
\begin{eqnarray*}
\mathfrak{S}^{\pm}_{1}&=&\lim_{t\to\pm\infty}\int_{\widehat{u}_{0}}^{+\infty}\,du\hspace{1mm}u^{\pm i\alpha^{'}-1}\int\limits_{\Omega_{III}}\,d\omega\hspace{1mm}\mathcal{F}(\omega)~e^{i\varphi^{\pm}_{(t)}(\omega,u)+i\delta(t)}=0,\\
\mathfrak{S}^{\pm}_{2}&=&\lim_{t\to\pm\infty}\int_{\widehat{u}_{0}}^{+\infty}\,du\hspace{1mm}u^{\pm i\alpha^{'}-1}\int\limits_{\Omega_{III}}\,d\omega\hspace{1mm}\mathcal{F}(\omega)~e^{i\widehat{\varphi}^{\pm}_{(t)}(\omega,u)+i\delta(t)}=0.
\end{eqnarray*}
\end{theorem}
\begin{proof}
We show the result for $\mathfrak{S}^{\pm}_{1}$, the proof for $\mathfrak{S}^{\pm}_{2}$ being similar. By introducing a convergence generating factor $e^{-\sigma u}$ we apply the Fubini theorem to obtain
\begin{equation}\label{sfrak}
\mathfrak{S}^{\pm}_{1}=\widehat{u}_{0}^{\pm i\alpha^{'}-1}\lim_{t\to\pm\infty}\lim_{\sigma\to 0^{+}}\int\limits_{\Omega_{III}}\,d\omega\hspace{1mm}\mathcal{F}(\omega)~e^{i(\omega-\omega^{'})t+i\delta(t)}\int_{1}^{+\infty}\,d\tau~\tau^{\pm i\alpha^{'}-1}~e^{-\widetilde{z}_{\pm}\tau}.
\end{equation}
The integral over $\tau$ can be computed as in Lemma~\ref{due}, and \eqref{sfrak} becomes
\begin{multline} \label{PARZ}
\mathfrak{S}^{\pm}_{1}=\pm\frac{i}{\alpha^{'}}\widehat{u}_{0}^{\pm i\alpha^{'}}\lim_{t\to\pm\infty}\lim_{\sigma\to 0^{+}}\int\limits_{\Omega_{III}}\,d\omega~\mathcal{F}(\omega)M(1,1\pm i\alpha^{'};\widetilde{z}_{\pm})~e^{i(\omega-\omega^{'})t+i\delta(t)-\widetilde{z}_{\pm}}\\
-\widehat{u}_{0}^{\pm i\alpha^{'}}\Gamma(\mp i\alpha^{'})\lim_{t\to\pm\infty}\lim_{\sigma\to 0^{+}}\int\limits_{\Omega_{III}}\,d\omega~\mathcal{F}(\omega)~\widetilde{z}_{\pm}^{\mp i\alpha^{'}}e^{i(\omega-\omega^{'})t+i\delta(t)}.
\end{multline}
Concerning the second term in the above expression we can immediately take the limit $\sigma\to 0^{+}$ inside the integral whereas for the first term in \eqref{PARZ} more care is required since the real part of the arguments entering in $M$ coincide. As a consequence 13.2.1 (ibid.) can not be used to find a $\sigma$-independent bound for $M(1,1\pm i\alpha^{'};\widetilde{z}_{\pm})$. However, $\mbox{Re}\widetilde{z}_{\pm}=\widehat{u}_{0}\sigma>0$, and 13.1.4 (ibid.) imply that for $C>0$ there exists a $\widetilde{\rho}$ such that
\begin{equation}\label{tilde}
|M(1,1\pm i\alpha^{'};z)|\leq|\Gamma(1\pm i\alpha^{'})|~e^{\widehat{u}_{0}\sigma}\left(1+\frac{C}{|z|}\right)
\end{equation}
for every $z\in\mathbb{C}\backslash K$ with $K:=\{z\in\mathbb{C}|~|\mbox{Re}z|\leq\widetilde{\rho},~|\mbox{Im}z|\leq\widetilde{\rho}~\}$. Notice that $M(1,1\pm i\alpha^{'};z)$ is bounded for every $z\in K$. Without loss of generality we can choose $C=1$, $\widetilde{\rho}>1$, and \eqref{tilde} gives the estimate $|M(1,1\pm i\alpha^{'};z)|\leq 2|\Gamma(1\pm i\alpha^{'})|~e^{\widehat{u}_{0}\sigma}$, from which it follows that $\left|e^{-\widetilde{z}_{\pm}}M(1,1\pm i\alpha^{'};\widetilde{z}_{\pm})\right|\leq  2|\Gamma(1\pm i\alpha^{'})|$. Hence, by applying the Lebesgue dominated convergence theorem \eqref{PARZ} simplifies to
\begin{multline*}
\mathfrak{S}^{\pm}_{1}=\pm\frac{i}{\alpha^{'}}\widehat{u}_{0}^{\pm i\alpha^{'}}\lim_{t\to\pm\infty}\int\limits_{\Omega_{III}}\,d\omega~\mathcal{F}(\omega)M(1,1\pm i\alpha^{'};z_{\pm})~e^{i(\omega-\omega^{'})t+i\delta(t)-z_{\pm}}\\
-(\mp i\widehat{u}_{0})^{\mp i\alpha^{'}}\Gamma(\mp i\alpha^{'})\lim_{t\to\pm\infty}\int\limits_{\Omega_{III}}\,d\omega~\mathcal{F}(\omega)(\kappa^{'}-\kappa)^{\mp i\alpha^{'}}e^{i(\omega-\omega^{'})t+i\delta(t)}
\end{multline*}
with $z_{\pm}=\mp i(\kappa^{'}-\kappa)\widehat{u}_{0}$. Let us define
\[
\widehat{\mathfrak{S}}^{\pm}_{1}:=\lim_{t\to\pm\infty}\int\limits_{\Omega_{III}}\,d\omega~\mathcal{F}(\omega)M(1,1\pm i\alpha^{'};z_{\pm})~e^{i(\omega-\omega^{'})t+i\delta(t)-z_{\pm}}.
\]
By adding, and subtracting a term $e^{i\widetilde{\delta}(t)}$ to $e^{i\delta(t)}$ we obtain
\begin{multline}\label{alphaalpha}
\widehat{\mathfrak{S}}^{\pm}_{1}=\lim_{t\to\pm\infty}\int\limits_{\Omega_{III}}\,d\omega~\mathcal{F}_{(t)}(\omega)M(1,1\pm i\alpha^{'};z_{\pm})~e^{i(\omega-\omega^{'})t-z_{\pm}}\\
+\lim_{t\to\pm\infty}e^{i\widetilde{\delta}(t)}\int\limits_{\Omega_{III}}\,d\omega~\mathcal{F}(\omega)M(1,1\pm i\alpha^{'};z_{\pm})~e^{i(\omega-\omega^{'})t-z_{\pm}}
\end{multline}
with $\mathcal{F}_{(t)}(\omega)$ defined as in Lemma~\ref{TEOREMA2}. Concerning the first term on the r.h.s. of \eqref{alphaalpha} we use \eqref{IDENTITA1} to integrate by parts, and we end up with
\[
\Sigma_{I}=-\lim_{t\to\pm\infty}\frac{1}{it}\int\limits_{\Omega_{III}}\,d\omega\frac{d}{d\omega}\left(\mathcal{F}_{(t)}(\omega)M(1,1\pm i\alpha^{'};z_{\pm})~e^{-z_{\pm}}\right)e^{i(\omega-\omega^{'})t}.
\]
According to 13.4.12 (ibid.) we have
\[
\frac{d}{d\omega}M(1,1\pm i\alpha^{'};z_{\pm})=\pm(i\widehat{u}_{0})^{-1}\left(\frac{\kappa}{\omega}\right)\left[M(1,1\pm i\alpha^{'};z_{\pm})\mp\frac{i\alpha^{'}}{1\pm i\alpha^{'}}M(1,2\pm i\alpha^{'};z_{\pm})\right]
\]
which is bounded since $M(1,1\pm i\alpha^{'};z_{\pm})$, and $M(1,2\pm i\alpha^{'};z_{\pm})$are bounded. Hence, we can compute $\Sigma_{I}$ by the same method used to evaluate \eqref{neviga3} in Lemma~\ref{TEOREMA2} and we conclude that $\Sigma_{I}=0$. Concerning the second term on the r.h.s. of \eqref{alphaalpha}, we apply the Riemann-Lebesgue lemma. Hence, $\widehat{\mathfrak{S}}^{\pm}_{1}=0$. Let us define
\[
\widehat{\mathfrak{S}}^{\pm}_{2}:=\lim_{t\to\pm\infty}\int\limits_{\Omega_{III}}\,d\omega~\mathcal{F}(\omega)(\kappa^{'}-\kappa)^{\mp i\alpha^{'}}~e^{i(\omega-\omega^{'})t+i\delta(t)}.
\]
After addition, and subtraction of $e^{i\widetilde{\delta}(t)}$ to $e^{i\delta(t)}$ we obtain
\begin{multline*}
\widehat{\mathfrak{S}}^{\pm}_{2}=\lim_{t\to\pm\infty}\int\limits_{\Omega_{III}}\,d\omega~\mathcal{F}_{(t)}(\omega)(\kappa^{'}-\kappa)^{\mp i\alpha^{'}}~e^{i(\omega-\omega^{'})t}\\
+\lim_{t\to\pm\infty}e^{i\widetilde{\delta}(t)}\int\limits_{\Omega_{III}}\,d\omega~\mathcal{F}(\omega)(\kappa^{'}-\kappa)^{\mp i\alpha^{'}}~e^{i(\omega-\omega^{'})t}.
\end{multline*}
The second term on the r.h.s. of the above expression is zero according to the Riemann-Lebesgue lemma. Concerning the other term we use \eqref{IDENTITA1} to integrate by parts, and we get
\[
\widehat{\mathfrak{S}}^{\pm}_{2}=-\lim_{t\to\pm\infty}\frac{1}{it}\int\limits_{\Omega_{III}}\,d\omega\frac{d}{d\omega}\left(\mathcal{F}_{(t)}(\omega)(\kappa^{'}-\kappa)^{\mp i\alpha^{'}}\right)e^{i(\omega-\omega^{'})t}
\]
which can be treated by means of the same methods used in Lemma~\ref{TEOREMA2} to compute the first integral entering in \eqref{TT1svil}. Hence, $\widehat{\mathfrak{S}}^{\pm}_{2}=0$.\hspace{5mm}$\square$ 
\end{proof}

\begin{theorem} \label{padrice}
Let $h(\omega,u)=\mathcal{O}(u^{-2})$ with $\partial_{\omega}\mathcal{O}(u^{-2})=\mathcal{O}(u^{-2})$, and $w\in\mathcal{C}^{1}(\Omega_{III})$ such that $|h(\omega,u)|\leq w(\omega)/u^{2}$, and $|w^{'}(\omega)|\leq C(1+\omega^{2})^{n}$ for some constants $C$, $n>0$. Then
\begin{eqnarray*}
\mathfrak{W}_{1}^{\pm}&=&\lim_{t\to\pm\infty}\int_{\widehat{u}_{0}}^{+\infty}\,du\hspace{1mm}u^{\pm i\alpha^{'}}\int\limits_{\Omega_{III}}\,d\omega\hspace{1mm}\mathcal{F}(\omega)h(\omega,u)~e^{i\varphi_{(t)}^{\pm}(\omega,u)+i\delta(t)}=0,\\
\mathfrak{W}_{2}^{\pm}&=&\lim_{t\to\pm\infty}\int_{\widehat{u}_{0}}^{+\infty}\,du\hspace{1mm}u^{\pm i\alpha^{'}}\int\limits_{\Omega_{III}}\,d\omega\hspace{1mm}\mathcal{F}(\omega)h(\omega,u)~e^{i\widehat{\varphi}_{(t)}^{\pm}(\omega,u)+i\delta(t)}=0
\end{eqnarray*}
with $\varphi_{(t)}^{\pm}(\omega,u)$, and $\widehat{\varphi}_{(t)}^{\pm}(\omega,u)$ as in Theorem\ref{TEOREMA3}.
\end{theorem}
\begin{proof}
We show $\mathfrak{W}_{1}^{\pm}$, the proof for $\mathfrak{W}_{2}^{\pm}$ being similar. Since $h(u,\omega)=\mathcal{O}(u^{-2})$ we can immediately apply the Fubini theorem, and we obtain
\begin{equation} \label{sonnig}
\mathfrak{W}_{1}^{\pm}=\lim_{t\to\pm\infty}\int\limits_{\Omega_{III}}\,d\omega~\mathcal{F}(\omega)~e^{i\delta(t)}N(\omega)e^{i(\omega-\omega^{'})t},\quad
N(\omega)=\int_{\widehat{u}_{0}}^{+\infty}\,du\hspace{1mm}u^{\pm i\alpha^{'}}h(\omega,u)~e^{\pm i(\kappa^{'}-\kappa)u}
\end{equation}
with $\left|N(\omega)\right|\leq w(\omega)/\widehat{u}_{0}$. By adding, and subtracting the term $e^{i\widetilde{\delta}(t)}$ to $e^{i\delta(t)}$, and applying the Riemann-Lebesgue lemma, \eqref{sonnig} reduces to
\[
\mathfrak{W}_{1}^{\pm}=\lim_{t\to\pm\infty}\int\limits_{\Omega_{III}}\,d\omega\hspace{1mm}\mathcal{F}_{(t)}(\omega)N(\omega)~e^{i(\omega-\omega^{'})t}.
\]
Finally, $\mathfrak{W}_{1}^{\pm}=0$ follows by using the same method employed in Lemma~\ref{TEOREMA2} to treat \eqref{referenza1} on the interval I. \hspace{5mm}$\square$
\end{proof}

\begin{theorem} \label{finalcut}
Let 
\[
\widetilde{Z}(\omega,u):=\left\{\begin{array}{ll}
                                                       1\\
                                                       u^{-1}\\
                                                       h(\omega,u)
                                                \end{array}\right\}\eta_{\infty}(u)
\]
with $h(\omega,u)$ as in Theorem~\ref{padrice}, and $\eta_{\infty}(u)$ as in Section~\ref{sec:3}. Then
\begin{eqnarray*}
\mathfrak{L}_{1}^{\pm}&=&\lim_{t\to\pm\infty}\int^{\widehat{u}_{0}}_{u_{0}}\,du\hspace{1mm}u^{\pm i\alpha^{'}}\eta_{\infty}(u)\int\limits_{\Omega_{III}}\,d\omega\hspace{1mm}\mathcal{F}(\omega)\widetilde{Z}(\omega,u)~e^{i\varphi_{(t)}^{\pm}(\omega,u)+i\delta(t)}=0,\\
\mathfrak{L}_{2}^{\pm}&=&\lim_{t\to\pm\infty}\int^{\widehat{u}_{0}}_{u_{0}}\,du\hspace{1mm}u^{\pm i\alpha^{'}}\eta_{\infty}(u)\int\limits_{\Omega_{III}}\,d\omega\hspace{1mm}\mathcal{F}(\omega)\widetilde{Z}(\omega,u)~e^{i\widehat{\varphi}_{(t)}^{\pm}(\omega,u)+i\delta(t)}=0.
\end{eqnarray*}
\end{theorem}
\begin{proof}
We show $\mathfrak{L}_{1}^{\pm}$, the proof for $\mathfrak{L}_{2}^{\pm}$ being similar. By applying the Fubini theorem we obtain
\begin{equation}\label{sonnig1}
\mathfrak{L}_{1}^{\pm}=\lim_{t\to\pm\infty}\int\limits_{\Omega_{III}}\,d\omega~\mathcal{F}(\omega)~e^{i\delta(t)}Q(\omega)e^{i(\omega-\omega^{'})t},\quad
Q(\omega)=\int^{\widehat{u}_{0}}_{u_{0}}\,du\hspace{1mm}u^{\pm i\alpha^{'}}\eta_{\infty}(u)\widetilde{Z}(\omega,u)~e^{\pm i(\kappa^{'}-\kappa)u}
\end{equation}
with
\[
\left|Q(\omega)\right|\leq(\widehat{u}_{0}-u_{0})\sup_{u\in[\widehat{u}_{0},u_{0}]}{\{|\widetilde{Z}|\}}=(\widehat{u}_{0}-u_{0})\left\{\begin{array}{ll}
                                                       1\\
                                                       u_{0}^{-1}\\
                                                       w(\omega)/u_{0}^{2}
                                                \end{array}\right\}.
\] 
By adding, and subtracting the term $e^{i\widetilde{\delta}(t)}$ to $e^{i\delta(t)}$, and applying the Riemann-Lebesgue lemma, \eqref{sonnig1} reduces to
\[
\mathfrak{L}_{1}^{\pm}=\lim_{t\to\pm\infty}\int\limits_{\Omega_{III}}\,d\omega\hspace{1mm}\mathcal{F}_{(t)}(\omega)Q(\omega)~e^{i(\omega-\omega^{'})t}.
\]
Finally, $\mathfrak{L}_{1}^{\pm}=0$ follows by using the same method adopted in Lemma~\ref{TEOREMA2} to treat \eqref{referenza1} on the interval I. \hspace{5mm}$\square$
\end{proof}
\begin{remark}
\em{Let us consider the definition of the modified wave operator \eqref{modificato} with $\mathcal{I}_{\infty}$ replaced by $\mathcal{I}_{\infty}-\mathcal{I}^{'}_{\infty}$, i.e.
\begin{equation}\label{mmmodificato}
s-\lim_{t \to \pm \infty}e^{-iHt}(\mathcal{I}_{\infty}-\mathcal{I}^{'}_{\infty})e^{iH_{\infty}t}\psi_{0}^{(\infty)}
\end{equation}
 where $\mathcal{I}^{'}_{\infty}$ is such that $\mathcal{I}_{\infty}-\mathcal{I}^{'}_{\infty}$ defines a new identifying operator with cut-off function $\chi$ having compact support $\mathscr{K}:=[u_{0},u_{0}^{'}]$, and $u_{0}^{'}>\widehat{u}_{0}>u_{0}>0$. Proceeding as in Section~\ref{sec:3} we can reduce \eqref{mmmodificato} to the computation of the following integrals
\begin{eqnarray*}
&&\lim_{t\to\pm\infty}\int\limits_{\mathscr{K}}\,du\hspace{1mm}u^{\pm i\alpha^{'}}\chi(u)\int\limits_{\Omega_{III}}\,d\omega\hspace{1mm}\mathcal{F}(\omega)\widetilde{Z}(u,\omega)~e^{i\varphi_{(t)}^{\pm}(\omega,u)+i\delta(t)},\\
&&\lim_{t\to\pm\infty}\int\limits_{\mathscr{K}}\,du\hspace{1mm}u^{\pm i\alpha^{'}}\chi(u)\int\limits_{\Omega_{III}}\,d\omega\hspace{1mm}\mathcal{F}(\omega)\widetilde{Z}(u,\omega)~e^{i\widehat{\varphi}_{(t)}^{\pm}(\omega,u)+i\delta(t)}.
\end{eqnarray*}
By applying the same method used to prove Theorem~\ref{finalcut} it can be showed that the above expressions are zero. Hence, \eqref{mmmodificato} is zero, implying that our definition \eqref{modificato} does not depend on the particular choice of $\mathcal{I}_{\infty}$.
}
\end{remark}
\begin{theorem}\label{omegaII}
Let $\omega^{'}\in\Omega_{II}$, $\beta=\sqrt{m_{e}^{2}-{\omega^{'}}^{2}}$, $\widetilde{\alpha}=Mm^{2}_{e}/\sqrt{m_{e}^{2}-{\omega^{'}}^{2}}$, and
\[
\widehat{Z}(\omega,u):=\left\{\begin{array}{ll}
                                                       1\\
                                                       u^{-1}\\
                                                       h(\omega,u)
                                                \end{array}\right\}
\]
with $h(\omega,u)$ as in Theorem~\ref{padrice}. Then
\begin{eqnarray}
\mathfrak{Z}_{1}^{\pm}&=&\lim_{t\to\pm\infty}\int_{\widehat{u}_{0}}^{+\infty}\,du\hspace{1mm}~e^{-\beta u}u^{\widetilde{\alpha}}\int\limits_{\Omega_{III}}\,d\omega\hspace{1mm}\mathcal{F}(\omega)\widehat{Z}(\omega,u)~e^{i(\omega-\omega^{'})t\pm i\kappa u+i\delta(t)}=0,\label{XXI}\\
\mathfrak{Z}_{2}^{\pm}&=&\lim_{t\to\pm\infty}\int^{\widehat{u}_{0}}_{u_{0}}\,du\hspace{1mm}\eta_{\infty}(u)e^{-\beta u}u^{\widetilde{\alpha}}\int\limits_{\Omega_{III}}\,d\omega\hspace{1mm}\mathcal{F}(\omega)\widehat{Z}(\omega,u)~e^{i(\omega-\omega^{'})t\pm i\kappa u+i\delta(t)}=0,\label{XXII}
\end{eqnarray}
\end{theorem}
\begin{proof}
By applying the Fubini theorem we obtain
\[
\mathfrak{Z}_{1}^{\pm}=\lim_{t\to\pm\infty}\int\limits_{\Omega_{III}}\,d\omega~\mathcal{F}(\omega)~e^{i\delta(t)}\widehat{Q}(\omega)e^{i(\omega-\omega^{'})t},\quad
\widehat{Q}(\omega)=\int_{\widehat{u}_{0}}^{+\infty}\,du\hspace{1mm}e^{-\beta u}u^{\widetilde{\alpha}}\widehat{Z}(\omega,u)~e^{\pm i\kappa u}
\]
with
\[
\left|Q(\omega)\right|\leq\int_{\widehat{u}_{0}}^{+\infty}\,du\hspace{1mm}e^{-\beta u}
 \left\{\begin{array}{ll}
                                                       u^{\widetilde{\alpha}}\\
                                                       u^{\widetilde{\alpha}-1}\\
                                                       w(\omega)u^{\widetilde{\alpha}-2}
                                                \end{array}\right\}=\left\{\begin{array}{ll}
 (e^{-x_{0}}x_{0}^{\widetilde{\alpha}}+\widetilde{\alpha}~\Gamma(\widetilde{\alpha},x_{0}))\beta^{-\widetilde{\alpha}-1}\\
 \Gamma(\widetilde{\alpha},x_{0})\beta^{-\widetilde{\alpha}}   \\
w(\omega)\Gamma(\widetilde{\alpha}-1,x_{0})\beta^{-\widetilde{\alpha}+1}
                                                \end{array}\right\}
\] 
where $x_{0}=\beta\widehat{u}_{0}>0$, and $\Gamma(\cdot,\cdot)$ is the incomplete Gamma function (see Erd$\acute{e}$ly et al., 9.1.2 Vol.II, p.136). Notice that for $\widetilde{\alpha}=1$ the incomplete Gamma function gives rise to an exponential integral $E_{1}(x_{0})=\Gamma(0,x_{0})$ which is well defined since $x_{0}>0$. By adding, and subtracting the term $e^{i\widetilde{\delta}(t)}$ to $e^{i\delta(t)}$, and applying the Riemann-Lebesgue lemma, we get
\[
\mathfrak{Z}_{1}^{\pm}=\lim_{t\to\pm\infty}\int\limits_{\Omega_{III}}\,d\omega\hspace{1mm}\mathcal{F}_{(t)}(\omega)\widehat{Q}(\omega)~e^{i(\omega-\omega^{'})t}.
\]
\eqref{XXI} follows by means of the same method used in Lemma~\ref{TEOREMA2} to treat \eqref{referenza1} whereas \eqref{XXII} can be obtained by proceeding as in Theorem~\ref{finalcut}.\hspace{5mm}$\square$
\end{proof}

\end{document}